# Ultra-Thin Lubricant-Infused Vertical Graphene Nanoscaffolds for High-Performance Dropwise Condensation

Abinash Tripathy[1,+], Cheuk Wing Edmond Lam[1,+], Diana Davila[2], Matteo Donati[1], Athanasios Milionis[1], Chander Shekhar Sharma[3], and Dimos Poulikakos[1*]

[1]Laboratory of Thermodynamics in Emerging Technologies, Department of Mechanical and Process Engineering, ETH Zurich, Sonneggstrasse 3, 8092 Zurich, Switzerland.

[2]IBM Research, Saeumerstrasse 4, 8803 Rueschlikon, Switzerland.

[3]Department of Mechanical Engineering, Indian Institute of Technology, Ropar, Nangal Road, Rupnagar, 140001 Punjab, India.

[*]Corresponding Author

Prof. Dr. Dimos Poulikakos
Email: dpoulikakos@ethz.ch
Phone: +41 44 632 27 38
Fax: +41 44 632 11 76

[+] These authors have contributed equally to this work.

**Abstract**

Lubricant-infused surfaces (LIS) are highly efficient in repelling water and constitute a very promising family of materials for condensation processes occurring in a broad range of energy applications. However, the performance of LIS in such processes is limited by the inherent thermal resistance imposed by the thickness of the lubricant and supporting surface structure, as well as by the gradual depletion of the lubricant over time. Here we present a remarkable, ultra-thin (~70 nm) and conductive LIS architecture, obtained by infusing lubricant into a vertically grown graphene nanoscaffold on copper. The ultra-thin nature of the scaffold, combined with the high in-plane thermal conductivity of graphene, drastically minimize earlier limitations, effectively doubling the heat transfer performance compared to a state-of-the-art CuO LIS surface. We show that the effect of the thermal resistance to the heat transfer performance of a LIS surface, although often overlooked, can be so detrimental that a simple nanostructured CuO surface can outperform a CuO LIS surface, despite filmwise condensation on the former. The present vertical graphene LIS is also found to be resistant to lubricant depletion, maintaining stable dropwise condensation for at least ~7 hours with no significant change of advancing contact angle and contact angle hysteresis. The lubricant consumed by the vertical graphene LIS is 52.6% less than the existing state-of-the-art CuO LIS, making also the fabrication process more economical.

**Introduction**

The phenomenon of condensation has many significant applications in technology of daily need, such as thermal management in high-performance computing, power generation, refrigeration, air conditioning and water desalination[1–4]. Its manifestation on a surface can be classified into two different modes depending on the condensate temporal accumulation and removal, i.e., filmwise condensation (FWC) and dropwise condensation (DWC)[5–7]. Due to the much higher heat transfer coefficient (HTC) by up to one order of magnitude for DWC compared to FWC, attaining and sustaining the dropwise mode is preferred for all applications where efficiency in energy transport is relevant[8–12].

Recently, employing lubricant-infused surfaces (LIS) in condensation heat transfer is justifiably receiving increased consideration. Unlike superhydrophobic surfaces, LIS are not prone to flooding of micro- and nanostructures (typical of more common lotus-based superhydrophobic surfaces[13–15]) during condensation at high supersaturations[16]. In the case of superhydrophobic surfaces with hierarchical textures, the extra surface area from the textures allows droplets to nucleate within them and often result in the formation of larger droplets in the Wenzel state on top of such structures, which adhere to the surface and degrade the heat transfer performance leading to condensate flooding and film formation[17,18]. For LIS, a lubricant with low surface energy and low vapor pressure is typically infused into the micro- and nanostructures present on the surface[19]. Capillary forces hold the lubricant within the structures and create an extremely smooth and chemically homogenous surface for condensation. This results in higher mobility of droplets on LIS, leading to easier continuous removal of condensate and more frequent regeneration of available nucleation sites, both critical to enable and enhance heat transfer by DWC[20–23]. Typically, the height of the porous structures infused by lubricant in LIS measures a few microns[15,20,21,24–26] and the amount of lubricant used for creating LIS is proportional to the height of these structures (Supplementary Table 1). In addition, most LIS have been fabricated on porous materials with low thermal conductivity such as polypyrrole on aluminum[27], surface modified and unmodified copper oxide (CuO)[20,21], paper[28]/cellulose[29], tungsten oxide on steel[30], nanoparticles coating[31] and porous polymer surfaces[32–35]. The thermal conductivity of lubricant used in LIS is also low (< 0.2 W m$^{-1}$ K$^{-1}$ for Krytox oils), which, together with the thickness and the relatively low conductivity of the porous materials employed, translates to an inherent significant thermal resistance of the lubricant layer, thus limiting the potential of using these surfaces for heat transfer applications.

Here, we have carefully designed an alternative LIS scaffold architecture, with unique attributes, which can alleviate previously mentioned deficiencies of the state-of-the-art. We report an ultrathin (~70 nm) LIS, obtained by infusing lubricant into a supporting vertically grown graphene nanoscaffold grown on copper, aiming at significantly promoting DWC and enhancing heat transfer.

Graphene is generally known to have very high in-plane thermal conductivity (2000 – 4000 W m$^{-1}$ K$^{-1}$)[36], which is attributed to the strong covalent sp$^2$ bonding between the carbon atoms (Supplementary Figure 1), whereas out-of-plane heat flow is limited by weak interlayer van der Waals coupling[36]. Between the in-plane and out-of-plane directions, graphene has therefore very high anisotropy in its thermal properties. In addition, graphene exhibits limited wettability close to the hydrophobic regime[37–39] and its wetting property can be tuned, for example, by changing the surface roughness[40,41].

Vertical graphene (VG), on the other hand, represents a special class of graphitic network, with its plates oriented perpendicularly to a substrate. The rising interest in VG as compared to planar horizontally oriented graphene stems from its vertical orientation itself, its high thermal conductivity in the vertical direction, its porosity and exposed sharp edges, its high surface-to-volume ratio and its non-stacking morphology[42–44]. VG also exhibits hydrophobic behavior[45] and the degree of hydrophobicity is higher than that of planar graphene[46] due to the presence of surface nanoroughness.

Porosity and high surface-to-volume ratio of VG indicate its potential for enhanced retention of lubricant by capillarity. Hence, we exploit these properties of VG to support infused lubricants and, what is critically important when minimal thermal resistance is targeted, we realize the fabrication of an ultrathin (~70 nm) VG nanoscaffold, constructing a LIS with outstanding performance with respect to condensation applications. The vertical orientation of graphene brings with it the additional heat transfer benefit related to its extreme thermal conductivity in the vertical direction (Figure 1a)[47]. As an additional benefit, the small thickness of VG LIS minimizes the amount of lubricant required to realize lubricant-infused texture compared to other LIS.

We have characterized the morphology, chemical composition and wettability of the above-described VG LIS using scanning electron microscopy (SEM), Raman spectroscopy and contact angle goniometry respectively. CuO LIS[48] and superhydrophilic CuO nanostructured surface[12] were chosen as state-of-the-art DWC and FWC reference surfaces, respectively, to compare the performance of VG LIS. Systematic heat transfer measurements at industrially relevant low-pressure steam conditions showed an approximately two-fold HTC increase with the VG LIS as compared to CuO LIS and superhydrophilic CuO nanostructured surfaces. We also performed prolonged condensation experiments for 7 hours to test the durability of the VG LIS as compared to CuO LIS. VG LIS maintained a twice as high HTC compared to CuO LIS throughout the durability experiment with minimal change in advancing contact angle and contact angle hysteresis, indicating sustainable droplet mobility on a surface.

## Results

**Fabrication of Ultra-Thin Lubricant-Infused Vertical Graphene Nanoscaffold Surfaces**

The growth of VG on copper was carried out using a thermal chemical vapor deposition (CVD) furnace. A catalyst-free approach was employed and $CH_4$ gas was used as the source of carbon. The time of growth of VG was varied between 15 and 60 minutes. Growth of VG for different time intervals was performed to study the temporal evolution of VG on copper. Growth time of 60 minutes was found to attribute sufficient uniformity and density to the grown VG nanoscaffold, used to impregnate the lubricant (Supplementary Figure 2). This growth time was used in all experiments in this work. Before infusing lubricant into the VG nanoscaffold, a ~5 nm layer of gold was coated on top, followed by dipping the surface in an ethanolic thiol solution to minimize the surface energy. The gold coating bonds chemically with the thiol. Krytox 1525 was then infused into the nanoscaffold to fabricate the targeted surfaces (refer to Methods for further details). It was found that this process significantly improved the adhesion between the lubricant and the VG nanoscaffold, thus enhancing lubricant retention as we will show later on.

**Characterization of Lubricant-infused Vertical Graphene Nanoscaffold Surfaces**

In order to realize the minimal thickness of VG network, we selected the minimum VG growth time of 60 minutes that resulted in the most uniform surface coverage. The SEM image in Figure 1b shows the morphology of VG grown for 60 minutes, revealing an interconnected VG nanoscaffold network covering uniformly the copper substrate. The height of VG for a growth time of 60 minutes was found to be ∼70 nm (Figure 1c). The density and thickness of the sidewalls of the VG were also found to increase with the growth time (Supplementary Figure 2). There was an increase in the sidewall thickness from ~13 nm to ~32 nm when the growth time was increased from 15 to 60 minutes (Supplementary Table 2).

Raman spectroscopy is a powerful, non-destructive and fast technique to study the properties of graphene. This technique provides information regarding the orientation and number of layers, and the presence of defects and edges, etc.[47,49,50]. Figure 1d shows the Raman spectra of the planar and VG samples used in our study. We observed the G peak (~1594 cm$^{-1}$) and the 2D peak (~2727 cm$^{-1}$) for all the samples. These bands are characteristic to graphitic structures and confirm the presence of graphene sheets[51–53]. In addition to that, we observed the D (~1365 cm$^{-1}$) and D+D' (~2972 cm$^{-1}$) peaks for the substrates with VG[54]. In planar graphene, these bands (D and D+D') are absent. The presence of high-intensity D and D+D' peaks signify a larger number of sharp edges in our VG samples[55].

The surface wetting property of VG was determined before and after the impregnation of lubricant. Using the sessile drop method, the contact angle measurement was performed with 8 µL deionized water droplets. Before lubricant impregnation, the advancing contact angle (ACA) of VG was 123° ± 2° (Supplementary Table 2 and Supplementary Figure 3) with a contact angle hysteresis (CAH) of 66° ± 2.5°, indicating high water adhesion, which inhibits droplet movement. After lubricant impregnation (see Methods), the ACA (inset in Figure 1c) was 119° ± 1° and CAH was 2° ± 1°. The massive decrease in CAH and the corresponding increased mobility of water droplets is critical for enhancing heat transfer through frequent removal of droplets.

Contact angle measurements were performed for the other two reference surfaces as well. The static contact angle (SCA) value for a CuO nanostructured surface (Supplementary Figure 4) was ~0° and the ACA and CAH values for CuO LIS were 119.7° ± 0.6° and 3.7° ± 1.5° respectively.

A CuO nanostructured surface was used as the reference for stable FWC[12] and CuO LIS was used to compare our surface with the existing state-of-the-art for DWC on lubricant-infused surfaces[21,48].

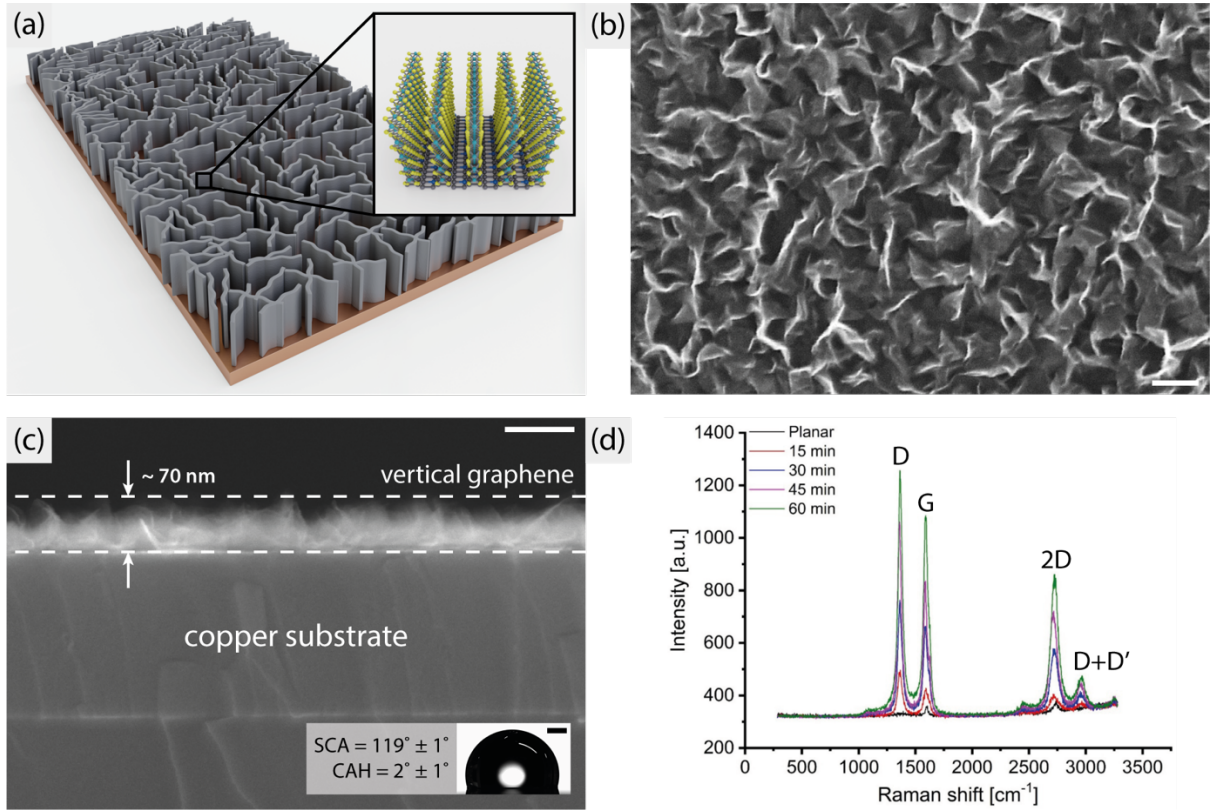

**Figure 1:** (a) Schematic of VG on copper. CVD was used to grow VG on copper using $CH_4$ gas as the carrier. (b) Top and (c) cross-sectional SEM images of the VG sample grown for 60 minutes (Scale bar – 100 nm). Inset: Image of water droplet (8 µL) on the VG LIS (Scale bar – 0.5 mm). (d) Raman spectroscopy of the planar graphene and VG samples with different growth times (15 min to 60 min). G and 2D bands are characteristic to graphitic structures and confirm the presence of graphene. D and D+D' bands were observed due to the presence of sharp edges on VG which were absent for the planar graphene sample.

To measure the amount of lubricant required to fabricate VG LIS and CuO LIS, the samples were weighed before and after infusing the lubricant. VG LIS (9 g m$^{-2}$) consumed 52.6% less lubricant than CuO LIS (19 g m$^{-2}$). This is due to the smaller height of VG (70 nm) as compared to CuO nanostructures (2 µm) and the larger spacing between the CuO nanostructures, which ranged from 0.3 – 0.7 µm, whereas the spacing between the sidewalls in the VG nanoscaffold was 30 – 90 nm. The smaller length scale of the vertical graphene nanoscaffold surface structures as compared to the CuO nanostructures indicates that the former will exhibit stronger capillary force to retain the lubricant.

The capillary length ($\kappa^{-1}$) of Krytox 1525 used in our study was ~1.01 mm ($\sqrt{\frac{\gamma}{\rho g}}$, $\gamma$ = 19 mN m$^{-1}$, $\rho$ = 1900 kg m$^{-3}$, $g$ = 9.8 m s$^{-2}$)[56]. The length scale of VG nanostructures is much smaller than the $\kappa^{-1}$ of Krytox 1525, which suggests its ability to enhance lubricant retention, benefiting from capillarity[57]. To study the lubricant retention ability of VG LIS further, we subjected the substrate to high shear conditions by spinning it at 5000 rpm (Supplementary Figure 5). In our experiment, the centrifugal force is the source of the high-shear condition. At this rotational speed, the acceleration in terms of g-force at a distance of 2 cm from the center of the substrate is $r\omega^2 = 0.02 \times (5000 \times 2\pi/60)^2 \approx 560g$, the location at which we measured the CAH on the substrates directly after the spinning test. The capillary length, $\kappa^{-1}$, of Krytox 1525 at $560g$ becomes 42.69 $\mu$m ($\kappa^{-1}$ is 21.34 $\mu$m at 10000 rpm). This value was still three orders of magnitude larger than the structures present on VG LIS, indicating that the surface should be able to retain the lubricant under such rapid spinning conditions. The measured CAH values on VG LIS were 2° ± 1° and 2.3° ± 0.6° before and after the spinning test, respectively, verifying this claim. The unchanged CAH after the spinning test indicated that VG LIS has superb lubricant retention ability under high-shear conditions, an important experimental design parameter to be considered while fabricating lubricant-infused surfaces, as the durability of the surface depends heavily on lubricant retention.

**Heat Transfer Coefficient Measurements**

In order to characterize the effect of reduced thermal resistance of VG LIS, we performed careful measurements of its heat transfer performance under industrially relevant operating conditions. We directly measured the condensation heat flux and condensation heat transfer coefficient and compared these to standard FWC heat transfer as well as DWC on state-of-the-art LIS. As discussed earlier, for FWC a CuO nanostructured surface was used as baseline reference. For DWC performance comparison CuO LIS was used.

The surfaces were placed vertically in a custom-built flow condensation chamber and were exposed to 50 mbar pure saturated steam flowing horizontally at ~3.7 m/s, while varying the sample temperature. Refer to Supplementary Sections 1 and 2 for details in experimental procedures and measurement, Supplementary Section 3 for chamber leakage estimation and Supplementary Section 4 for steam flow speed estimation.

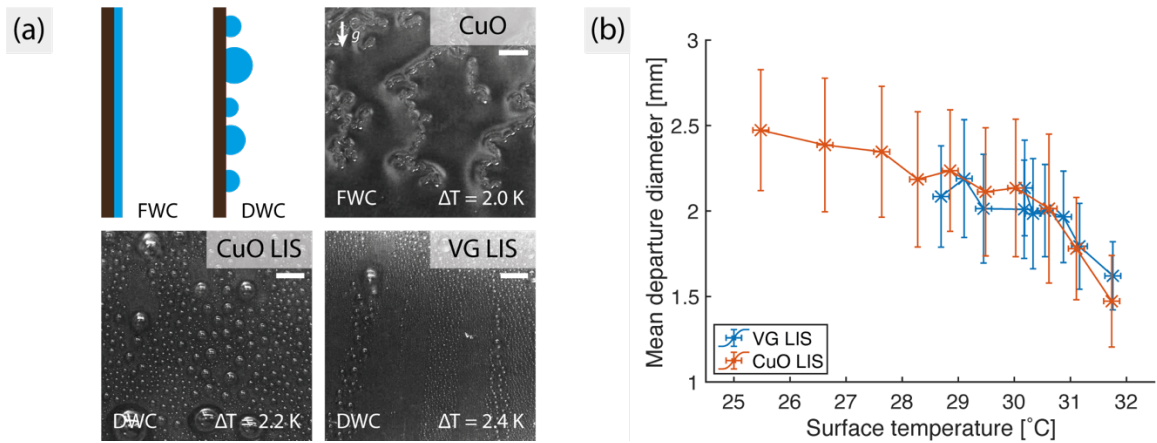

**Figure 2:** (a) Condensation modes on the CuO nanostructured surface, CuO LIS and VG LIS at ~ 2 K subcooling. Schematic shows a side view of the 2 condensation modes. A continuous film of water on the CuO nanostructured surface presents as a significant additional thermal resistance on top of the surface compared to periodically shedding condensate droplets for CuO LIS and VG LIS surfaces (Scale bar – 2 mm). (b) Mean departure diameter on the two surfaces as a function of surface temperature $T_{\mathrm{surf}}$. Steam temperature was kept constant at ~32.9°C. The trend of decreasing diameters with increasing surface temperatures was observed on both surfaces. Vertical

error bars indicate the standard deviation of the departure diameter distribution. Horizontal error bars indicate uncertainty in the surface temperature measurements.

Figure 2a shows visually the condensation behavior on the 3 surfaces at a subcooling value $\Delta T$ around 2 K. Videos of the condensation behavior on the 3 surfaces can be found in Supplementary Movie S1. It is clear that dropwise condensation manifested itself on the lubricant-infused surfaces of VG LIS and CuO LIS whereas the condensation was filmwise on the superhydrophilic CuO nanostructured surface. We measured departure diameters for all droplet departure events in a 1-minute period at different subcooling values. The mean measured diameters are presented in Figure 2b as a function of the surface temperature, $T_{\text{surf}}$. Error estimation can be found in Supplementary Section 6. Droplet measurement procedures can be found in Supplementary Section 7. The minimum attainable surface temperature was limited by the performance of the surface, i.e., a more efficient surface renders a smaller range of attainable surface temperatures as there is less resistance for the surface to approach the fixed steam temperature. The two surfaces exhibited similar diameters, and the same general trend of decreasing diameters with increasing surface temperatures, which we attribute to the increase of nucleation rate at lower surface temperatures, given the same steam temperature of ~32.9 ˚C. The departure droplet diameters span mainly between 1.5 and 2.5 mm, slightly below the capillary length of 2.7 mm for water. The full distribution of the measurements can be found in Supplementary Figure 6. The results suggest that the surface properties are similar during condensation for both surfaces, being dominated by the lubricant.

We then investigated the condensation behavior on the 3 surfaces as a function of the subcooling, defined as $\Delta T = T_{\text{steam}} - T_{\text{base}}$, where $T_{\text{steam}}$ is the steam temperature (~32.9 ˚C) and $T_{\text{base}}$ is the temperature below the lubricant layer computed from the Fourier's law of thermal conduction (Supplementary Section 5).

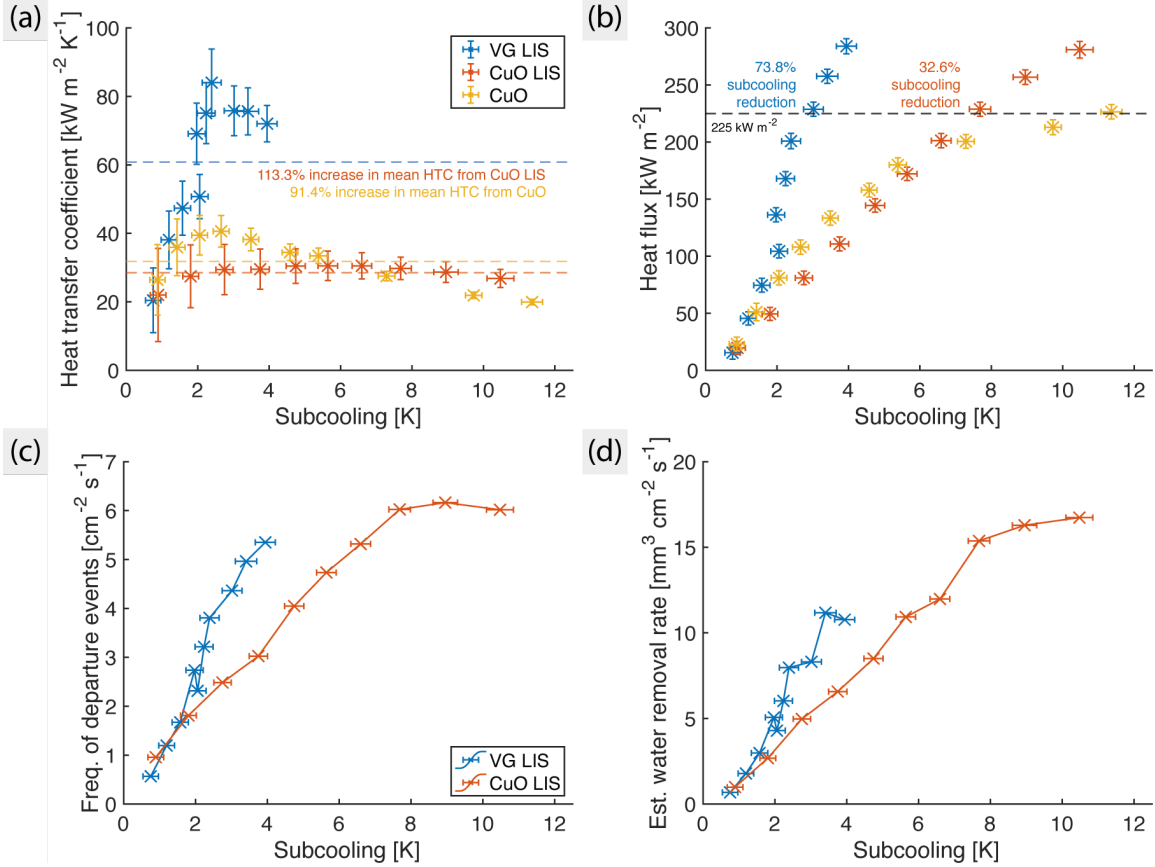

**Figure 3:** (a) HTC of VG LIS, CuO LIS, and the CuO nanostructured surface at different subcooling values $\Delta T = T_{\text{steam}} - T_{\text{base}}$. The range of the attained subcooling values is influenced by the performance of the interface. A better performing surface shifts the maximum attainable subcooling value to the left (toward smaller subcooling values). On average across these 10 data points (arithmetic mean), VG LIS presented a 113.3% increase from CuO LIS, and a 91.4% increase from the CuO nanostructured surface. (b) Corresponding heat flux measurements. At the same subcooling values, VG LIS can maintain the highest heat flux among the 3 surfaces. (c) Rate of droplet departure measured on VG LIS and CuO LIS in an area of 10 mm x 10 mm around the center of the sample. At 2 K subcooling and beyond, droplet departure was more frequent on VG LIS than CuO LIS. (d) Estimated water removal rate from the 2 surfaces based on measured individual droplet departure diameters. With similar mean departure diameters in Figure 2b and the deviation of droplet departure frequency of the 2 surfaces after 2 K subcooling (Figure 3c), the water removal rate after 2 K on VG LIS deviated from CuO LIS as well.

The HTC was computed from the measured sample temperature and heat flux into the cooled surface. We defined the HTC as $h = q''/\Delta T$, where $q''$ is the measured heat flux. The use of $\Delta T = T_{\text{steam}} - T_{\text{base}}$ computed from the base temperature below the lubricant layer naturally incorporated its thermal resistance into the HTC definition. In Figure 3a, we report the HTC measurements for each surface at 10 different subcooling values, obtained by setting the heat sink at 10 different temperatures for a constant saturated steam temperature at 50 mbar. As a better performing surface introduces less interfacial heat transfer resistance, the subcooling it yields will naturally drift to the lower values, at the same time resulting in higher HTCs. Computing the arithmetic mean across all values investigated, CuO LIS performed similarly to the superhydrophilic CuO nanostructured surface. In contrast, VG LIS performed 91.4% better on average compared to the CuO nanostructured surface and 113.3% better compared to CuO LIS. At the subcooling of 2.4 K where VG LIS performed the best, the improvement over the CuO nanostructured surface was 109.2% and that over CuO LIS was 192.4%. The benefit of DWC on CuO LIS was diminished by the thermal resistance accompanied with its relatively thick lubricant layer.

The corresponding heat flux measurements are shown in Figure 3b. At a heat flux of 225 kW m$^{-2}$, VG LIS could sustain the condensation at 73.8% smaller subcooling on the average than the CuO nanostructured surface, compared to a 32.6% smaller subcooling of CuO LIS. At 3.8 K subcooling, the heat flux yielded by VG LIS was 97.4% higher than that of the CuO nanostructured surface and 146.3% higher than that of CuO LIS.

These results indicate that considering the thermal resistance of the lubricant layer, which is inherent in all LIS, the improvement of HTC for DWC on CuO LIS was effectively lost. If this resistance was left out from the computation of the HTC, CuO LIS would show an artificial improvement over reference filmwise condensation on the CuO nanostructured surface. The apparent restoration, however, is solely a result of neglecting to account for the lubricant layer thermal resistance. For the sake of completeness, Supplementary Section 8 describes the measured HTCs when the thermal resistance is not taken into account.

We emphasize here the significant and detrimental impact on the heat transfer efficiency of using low-adhesion layers to achieve DWC which is often neglected in coating design. In attempts to achieve DWC with LIS, these layers add simultaneously thermal resistance due to the low thermal conductivity of the porous matrix as well of the lubricant, which are typically both poor thermal conductors. For lubricants, the thermal conductivity is typically on the order of 0.1 W m$^{-1}$ K$^{-1}$, three orders of magnitude lower compared to metals with thermal conductivities on the order of 100 W m$^{-1}$ K$^{-1}$. The porous matrix in typical LIS is composed of metal oxides which also have thermal conductivities far below that of metals, e.g., CuO[58] has a thermal conductivity of 33 W m$^{-1}$ K$^{-1}$. Therefore, to truly harness the benefits of DWC a low-adhesion lubricant-infused texture with simultaneously minimal thermal resistance is required and consequently, the effect of thermal resistance must be taken into account to correctly determine the heat transfer effectiveness of the lubricant-infused texture.

Our approach circumvents this inherent and prohibitive drawback of LIS from two very effective directions: By significantly reducing the thickness of the lubricant-infused texture, and by increasing the thermal conductivity of the solid porous matrix using a vertically oriented graphene nanoscaffold. Using the thermal conductivities of the lubricant Krytox 1525 (0.1 W m$^{-1}$ K$^{-1}$), vertical graphene[47] (250 W m$^{-1}$ K$^{-1}$) and CuO[58] (33 W m$^{-1}$ K$^{-1}$), as well as the heights of the structures, we computed that the thermal resistance of the lubricant-infused nanotextures to be 0.0011 K W$^{-1}$ for VG LIS and 0.0317 K W$^{-1}$ for CuO LIS for a 20 mm x 20 mm surface, the area exposed to steam in our condensation experiments. At a heat flux of 225 kW m$^{-2}$ in our experiments, for example, this would translate to a temperature difference of ~0.1 K across the lubricant layer for VG LIS, but ~2.9 K for CuO LIS. Normalized to a condensing surface area of 1 $m^2$, the thermal resistance values would be 4.53 x $10^{-7}$ K $W^{-1}$ and 1.27 x $10^{-5}$ K $W^{-1}$ for VG LIS and CuO LIS respectively. In other words, we achieved, by design, an order-of-magnitude reduction in thermal resistance of the lubricant layer in VG LIS compared to state-of-the-art CuO LIS which directly contributes to the difference in heat transfer performance illustrated in Figures 3a and 3b. Refer to Supplementary Section 5 for details on the estimation of thermal resistance of LIS.

The number of departure events shown in Figure 3c supports the HTC measurements. We chose a window of 10 mm x 10 mm around the center of the sample for measuring droplet departure (Supplementary Section 7) to avoid edge effects. With similar mean departure diameters in Figure 2b, a higher number of departure events at the same subcooling suggests more water condensed, and thus a higher condensation heat transfer efficiency. The droplet departure event curves for VG LIS and CuO LIS diverge markedly beyond 2 K subcooling, consistent with HTC and heat flux curves in Figures 3a and 3b. Similarly, we estimated the water removal rate of the 2 samples by using the individually measured departure diameters, assuming a spherical cap droplet shape, calculated based on the static contact angle for departing water droplets in Figure 3d. This is a rough estimation since CAH and the coalescence events of the departing droplets immediately before measurement may distort their shape from a perfect spherical cap geometry. Nevertheless, a clear deviation beyond 2 K was still observed.

In addition to the heat transfer performance of a LIS, it is also important to verify whether the texture can retain the lubricant successfully under condensation conditions over time, where the shear from sliding droplets and steam flow can cause a gradual drainage and loss of lubricant, and consequent degradation in surface heat transfer performance. We demonstrate in the following section that VG LIS can sustain its performance for long periods of time under industrially relevant conditions.

**Durability Test**

To test the durability performance of the VG LIS surface, we exposed it to the same conditions as in the HTC measurement experiments, i.e., saturated steam at 50 mbar, for 7 hours. To our knowledge, studies of this type, i.e., prolonged condensation of water on lubricated surfaces, to allow for comparison have not been performed in the literature, despite the clear need to address the performance of LIS over time.

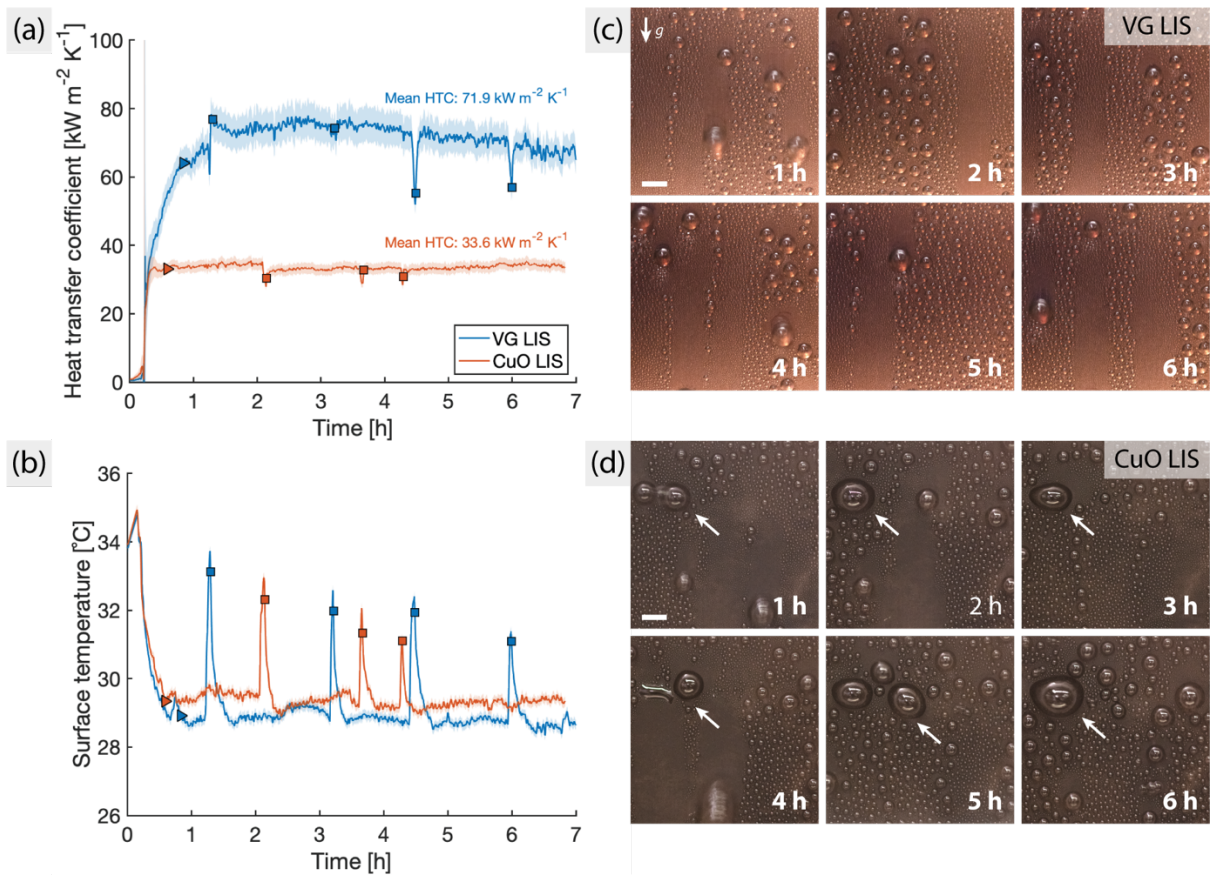

**Figure 4**: (a) HTC measurement on VG LIS and CuO LIS for ~ 7 h, as 1-minute moving average. $t = 0$ when steam was first introduced into the evacuated chamber. Triangles refer to the time points when the heat sink reached the target temperature. Squares denote when the flow was reduced temporarily for maintenance. (b) Corresponding surface temperature evolution during the 7-h period as 1-minute moving average. Appearance of condensate droplets on (c) VG LIS and (d) CuO LIS during the experiment. White arrows indicate pinning sites developed on CuO LIS (Scale bars – 2 mm).

The performance of VG LIS and CuO LIS are reported in Figure 4. We set $t = 0$ to be the time when steam was introduced into the evacuated chamber. When the chamber pressure reached the target pressure of 50 mbar, the samples were subsequently set to the target subcooling value by cooling the heat sink. We compared VG LIS and CuO LIS at similar surface temperature $T_{\text{surf}}$ which, at the same steam temperature, results in a similar nucleation rate, thus the shear dynamics on the lubricant. We targeted a surface temperature $T_{\text{surf}} \approx 29\ °\text{C}$ at the beginning of the durability test and the corresponding heat sink temperature was fixed for the entire 7 hours. In the 7-hour period, there were occasions when steam flow had to be reduced temporarily for a few minutes for maintenance of our setup, as indicated by squares in Figure 4a and 4b. These resulted in a temporary perturbation in measured parameters, but all returned to nominal or original values after the short maintenance period. For the entire duration of 7 h, VG LIS maintained a significantly higher HTC (114.1% average improvement in temporal mean HTC) than the CuO LIS reference, as shown in Figure 4a.

Photographs of both surfaces at different times are shown in Figures 4c and 4d. A pinning site (indicated by arrows) was seen early in the experiment on CuO LIS and developed as time progressed. After the durability test, CAH was measured on both the surfaces, and the increase as compared to prior the test is markedly more prominent on CuO LIS (+6.1°) than VG LIS (+2.7°). A time-lapse movie of the condensation on both surfaces can be found in Supplementary Movie S2.

**Conclusion**

We have introduced and investigated a novel lubricant-infused surface architecture based on VG nanoscaffolds grown on copper, which can exhibit stable DWC of water for at least ~7 hours, meanwhile achieving significant heat transfer improvements over the state-of-the-art CuO LIS. The key advance in the material design comes from the vertical, highly conductive and ultra-thin graphene nanostructures that are able to support a proportionally thin lubricant layer (~70 nm), the layer thinness being a critical advancement over the state-of-the-art, regarding heat transfer performance. This minimizes the inherent thermal resistance of this composite coating (order-of-magnitude reduction compared to the reference state-of-the-art CuO LIS), while exhibiting excellent lubricant retention (consuming only half of the lubricant as compared to CuO LIS) and yielding a 2-fold heat transfer efficiency. The work brings to the forefront the critical need to design LIS surfaces for heat transfer condensation applications with minimal thermal resistance and always rate them without neglecting their thermal resistance. The simple and scalable fabrication of the VG LIS coating reported here significantly advances the state-of-the-art and offers itself as a promising candidate for a host of condensation applications.

**Methods**

**Growth of Vertical Graphene (VG)**

Vertical graphene was grown using a catalyst-free method in a thermal CVD furnace (TCVD RF100CA 2G growth tool, Graphene Square Inc.) with a tubular design configuration and equipped with an inductively coupled plasma (ICP) unit. The substrates (copper pieces acquired from Metall Service Menziken, Switzerland, 1.5 mm thick, 5 cm x 2 cm) were cleaned in subsequent baths of acetone, isopropanol and DI water. $O_2$ plasma was then used to eliminate any possible organic contaminants on the sample. This step oxidized the surface of the copper, which was further etched for 1 minute in a mixture of HCl:H$_2$O (1:10) prior to growth.

The samples were then immediately introduced in the CVD furnace at room temperature and heated up to 750°C under a 20 sccm H$_2$ flow. Once this temperature was reached, the samples were allowed to stabilize for 10 minutes under the same H$_2$ flow. At this point, methane (CH$_4$) was introduced as the carbon source for the vertical growth of vertical graphene nanosheets. A H$_2$:CH$_4$ flow ratio of 6:9, a RF power of 200 W and a pressure of 0.045 torr were used. The growth time was varied from 15 to 60 minutes to achieve different layer thicknesses. Following the growth of the vertical graphene, the substrates were allowed to cool down under a 20 sccm H$_2$ flow until room temperature was reached. The samples were then removed from the furnace. No post-transfer procedure was needed afterwards as the vertical graphene grew directly on the surface of the copper samples. Only for 60 minutes growth time, the growth of VG structures was uniform all across the surface and dense. Hence, nanoscaffold VG samples grown for 60 minutes were used in all our experiments and the growth time was not increased further. The growth of VG structures on surfaces other than copper can be performed with an additional thin layer of copper (100-150 nm[59]) deposited on top of the surface of interest, making the above fabrication process versatile for a broad range of materials.

**Surface Characterization of VG**

The morphological structure of the vertical graphene was analyzed using a scanning electron transmission microscope (SU8230, Hitachi). A Raman spectroscopy setup (NT-MDT NTEGRA Spectra) was employed to verify the chemical composition of the samples. Raman measurements were carried out using a 50-mW solid-state laser with an excitation wavelength of 473.05 nm in single longitudinal mode and linearly polarized. The light was collected using a 100x long-working distance objective (working distance of 7 mm and numerical aperture of 0.7) and detected by a Peltier-cooled ultra-low noise charge-coupled device (CCD) camera. A 600 lines per mm grating was used to measure the complete Raman spectra at localized areas with an acquisition time of 10 seconds.

**Fabrication of Lubricant Infused VG**

To prepare the lubricant infused VG samples, first the samples were coated with 5 nm of gold using e-beam evaporation (BAK 501LL, Evatech). For practical applications, this can be replaced with an ultra-thin layer of either copper or aluminium. The gold coated VG samples were dipped into an ethanolic thiol solution for one hour[10] to minimize the surface energy prior to infusing the lubricant. Gold coating on VG was performed to achieve a strong chemical bond with thiol. Then the samples were dipped in the lubricant (Krytox 1525, Sigma Aldrich, dynamic viscosity 496 mPa s, surface tension 19 mN m$^{-1}$, density 1900 kg m$^{-3}$) for 15 minutes. After that the samples were kept vertically oriented for one day followed by $N_2$ purge to remove the excess lubricant. We determined the wettability of all the fabricated substrates by performing water contact angle and contact angle hysteresis measurement using a goniometer (OCA 35, DataPhysics). Water droplets of volume 8 $\mu L$ were used for the all measurements.

**Fabrication of CuO Nanostructured Superhydrophilic Surface and CuO Lubricant-Infused Surface**

The copper substrate was first cleaned in an ultrasonic bath with acetone, IPA and DI water followed by drying in $N_2$ blow. Then the substrate was dipped in to 2M hydrochloric acid (HCl) for 50-60 seconds to remove the oxide layer. The substrate was again rinsed with DI water and dried with $N_2$. To fabricate the CuO nanostructured surface, the cleaned copper substrate was dipped in a hot alkali solution (mixture of $NaClO_2$, NaOH, $Na_3PO_4:12H_2O$) for 4-5 minutes[60]. The temperature of the alkali solution was maintained at 95°C. The copper samples turned black after the reaction due to the formation of CuO nanoblades. The morphology of the surface was observed using SEM (SU8230, Hitachi) shown in Supplementary Figure 4. The height of CuO nanostructures formed was $h \approx 2~\mu m$ with a thermal conductivity of $20 - 33$ W m$^{-1}$ K$^{-1}$ [21,58,61]. To fabricate the CuO lubricant-infused surfaces, samples were first dipped in ethanolic thiol (Perfluordecanthiol (PFDT), Sigma Aldrich) solution followed by dipping in Krytox 1525 for 15 minutes and then the samples were kept vertically for 24 hours to drain the excess lubricant.

**Heat Transfer Measurement and Durability Test**

See Supplementary Sections 1 and 2 for details.

**Data Availability Statement**

All the data used in the main manuscript and SI to support the claims are available from the corresponding author upon reasonable request.

**Acknowledgements**

We would like to thank Mr. Jovo Vidic for his help in the setup of the data acquisition system and Mr. Peter Feusi for his help in the construction of the condensation chamber for the heat transfer experiments. We thank the Cleanroom Operations Team of the Binnig and Rohrer Nanotechnology Center (BRNC) for their help and support.

**Funding Information-** This project has received funding from the European Union's Horizon 2020 research and innovation programme under grant number 801229 (HARMoNIC).

**Author contributions**

A.T. and D.P. conceived the research. A.M., C.S.S. and D.P. provided scientific guidance in all of its aspects. A.T. and D.D.P. fabricated and characterized the samples. C.W.E.L. designed and constructed the condensation chamber; and performed all the experiments and data analyses related to condensation and heat transfer. M.D. applied the gold coating on vertical graphene. The manuscript was written with the contribution of all the authors.

[+] A.T. and C.W.E.L. have contributed equally to this work.

**References**

(1) Enright, R.; Miljkovic, N.; Alvarado, J. L.; Kim, K.; Rose, J. W. Dropwise Condensation on Micro-and Nanostructured Surfaces. *Nanoscale Microscale Thermophys. Eng.* **2014**, *18*, 223–250.

(2) Attinger, D.; Frankiewicz, C.; Betz, A. R.; Schutzius, T. M.; Ganguly, R.; Das, A.; Kim, C.-J.; Megaridis, C. M. Surface Engineering for Phase Change Heat Transfer: A Review. *MRS Energy Sustain.* **2014**, *1*, 1–40.

(3) Cho, H. J.; Preston, D. J.; Zhu, Y.; Wang, E. N. Nanoengineered Materials for Liquid-Vapour Phase-Change Heat Transfer. *Nat. Rev. Mater.* **2016**, *2*, 1–17.

(4) Edalatpour, M.; Liu, L.; Jacobi, A. M.; Eid, K. F.; Sommers, A. D. Managing Water on Heat Transfer Surfaces: A Critical Review of Techniques to Modify Surface Wettability for Applications with Condensation or Evaporation. *Appl. Energy* **2018**, *222*, 967–992.

(5) Paxson, A. T.; Yagüe, J. L.; Gleason, K. K.; Varanasi, K. K. Stable Dropwise Condensation for Enhancing Heat Transfer via the Initiated Chemical Vapor Deposition (ICVD) of Grafted Polymer Films. *Adv. Mater.* **2014**, *26*, 418–423.

(6) Rykaczewski, K.; Paxson, A. T.; Anand, S.; Chen, X.; Wang, Z.; Varanasi, K. K. Multimode Multidrop Serial Coalescence Effects during Condensation on Hierarchical Superhydrophobic Surfaces. *Langmuir* **2013**, *29*, 881–891.

(7) Mulroe, M. D.; Srijanto, B. R.; Ahmadi, S. F.; Collier, C. P.; Boreyko, J. B. Tuning Superhydrophobic Nanostructures to Enhance Jumping-Droplet Condensation. *ACS Nano*

**2017**, *11*, 8499–8510.

(8) Schmidt, E.; Schurig, W.; Sellschopp, W. Versuche Über Die Kondensation von Wasserdampf in Film- Und Tropfenform. *Tech. Mech. und Thermodyn.* **1930**, *1*, 53–63.

(9) Rose, J. W., Dropwise Condensation Theory and Experiment: A Review. *Proc. Inst. Mech. Eng. Part A J. Power Energy* **2002**, *216*, 115–128.

(10) Sharma, C. S.; Stamatopoulos, C.; Suter, R.; Von Rohr, P. R.; Poulikakos, D. Rationally 3D-Textured Copper Surfaces for Laplace Pressure Imbalance-Induced Enhancement in Dropwise Condensation. *ACS Appl. Mater. Interfaces* **2018**, *10*, 29127–29135.

(11) Sharma, C. S.; Combe, J.; Giger, M.; Emmerich, T.; Poulikakos, D. Growth Rates and Spontaneous Navigation of Condensate Droplets Through Randomly Structured Textures. *ACS Nano* **2017**, *11*, 1673–1682.

(12) Donati, M.; Lam, C. W. E.; Milionis, A.; Sharma, C. S.; Tripathy, A.; Zendeli, A.; Poulikakos, D. Sprayable Thin and Robust Carbon Nanofiber Composite Coating for Extreme Jumping Dropwise Condensation Performance. *Adv. Mater. Interfaces* **2020**, *8*, 2001176.

(13) Liu, Y.; Choi, C. H. Condensation-Induced Wetting State and Contact Angle Hysteresis on Superhydrophobic Lotus Leaves. *Colloid Polym. Sci.* **2013**, *291*, 437–445.

(14) Jo, H.; Hwang, K. W.; Kim, D.; Kiyofumi, M.; Park, H. S.; Kim, M. H.; Ahn, H. S. Loss of Superhydrophobicity of Hydrophobic Micro/Nano Structures during Condensation. *Sci. Rep.* **2015**, *5*, 5–10.

(15) Dai, X.; Sun, N.; Nielsen, S. O.; Stogin, B. B.; Wang, J.; Yang, S.; Wong, T. S. Hydrophilic Directional Slippery Rough Surfaces for Water Harvesting. *Sci. Adv.* **2018**, *4*, 1–11.

(16) Tsuchiya, H.; Tenjimbayashi, M.; Moriya, T.; Yoshikawa, R.; Sasaki, K.; Togasawa, R.; Yamazaki, T.; Manabe, K.; Shiratori, S. Liquid-Infused Smooth Surface for Improved Condensation Heat Transfer. *Langmuir* **2017**, *33*, 8950–8960.

(17) Miljkovic, N.; Enright, R.; Wang, E. N. Effect of Droplet Morphology on Growth Dynamics and Heat Transfer during Condensation on Superhydrophobic Nanostructured Surfaces. *Am. Chem. Soc.* **2012**, *6*, 1776–1785.

(18) Chen, X.; Wu, J.; Ma, R.; Hua, M.; Koratkar, N.; Yao, S.; Wang, Z. Nanograssed Micropyramidal Architectures for Continuous Dropwise Condensation. *Adv. Funct. Mater.* **2011**, *21*, 4617–4623.

(19) Baumli, P.; D'Acunzi, M.; Hegner, K. I.; Naga, A.; Wong, W. S. Y.; Butt, H. J.; Vollmer, D. The Challenge of Lubricant-Replenishment on Lubricant-Impregnated Surfaces. *Adv. Colloid Interface Sci.* **2021**, *287*, 102329.

(20) Xiao, R.; Miljkovic, N.; Enright, R.; Wang, E. N. Immersion Condensation on Oil-Infused Heterogeneous Surfaces for Enhanced Heat Transfer. *Sci. Rep.* **2013**, *3*, 1988.

(21) Sett, S.; Sokalski, P.; Boyina, K.; Li, L.; Rabbi, K. F.; Auby, H.; Foulkes, T.; Mahvi, A.; Barac, G.; Bolton, L. W.; Miljkovic, N. Stable Dropwise Condensation of Ethanol and Hexane on Rationally Designed Ultrascalable Nanostructured Lubricant-Infused Surfaces. *Nano Lett.* **2019**, *19*, 5287–5296.

(22) Sun, J.; Weisensee, P. B. Microdroplet Self-Propulsion during Dropwise Condensation on Lubricant-Infused Surfaces. *Soft Matter* **2019**, *15*, 4808–4817.

(23) Anand, S.; Paxson, A. T.; Dhiman, R.; Smith, J. D.; Varanasi, K. K. Enhanced Condensation on Lubricant-Impregnated Nanotextured Surfaces. *ACS Nano* **2012**, *6*, 10122–10129.

(24) Adera, S.; Alvarenga, J.; Shneidman, A. V.; Zhang, C. T.; Davitt, A.; Aizenberg, J. Depletion of Lubricant from Nanostructured Oil-Infused Surfaces by Pendant Condensate Droplets. *ACS Nano* **2020**, *14*, 8024–8035.

(25) Wong, T. S.; Kang, S. H.; Tang, S. K. Y.; Smythe, E. J.; Hatton, B. D.; Grinthal, A.; Aizenberg, J. Bioinspired Self-Repairing Slippery Surfaces with Pressure-Stable Omniphobicity. *Nature* **2011**, *477*, 443–447.

(26) Epstein, A. K.; Wong, T. S.; Belisle, R. A.; Boggs, E. M.; Aizenberg, J. Liquid-Infused Structured Surfaces with Exceptional Anti-Biofouling Performance. *Proc. Natl. Acad. Sci.* **2012**, *109*, 13182–13187.

(27) Kim, P.; Wong, T. S.; Alvarenga, J.; Kreder, M. J.; Adorno-Martinez, W. E.; Aizenberg, J. Liquid-Infused Nanostructured Surfaces with Extreme Anti-Ice and Anti-Frost Performance. *ACS Nano* **2012**, *6*, 6569–6577.

(28) Glavan, A. C.; Martinez, R. V.; Subramaniam, A. B.; Yoon, H. J.; Nunes, R. M. D.; Lange, H.; Thuo, M. M.; Whitesides, G. M. Omniphobic "RF Paper" Produced by Silanization of Paper with Fluoroalkyltrichlorosilanes. *Adv. Funct. Mater.* **2014**, *24*, 60–70.

(29) Guo, J.; Fang, W.; Welle, A.; Feng, W.; Filpponen, I.; Rojas, O. J.; Levkin, P. A. Superhydrophobic and Slippery Lubricant-Infused Flexible Transparent Nanocellulose Films by Photoinduced Thiol-Ene Functionalization. *ACS Appl. Mater. Interfaces* **2016**, *8*, 34115–34122.

(30) Tesler, A. B.; Kim, P.; Kolle, S.; Howell, C.; Ahanotu, O.; Aizenberg, J. Extremely Durable Biofouling-Resistant Metallic Surfaces Based on Electrodeposited Nanoporous Tungstite Films on Steel. *Nat. Commun.* **2015**, *6*, 8649.

(31) Juuti, P.; Haapanen, J.; Stenroos, C.; Niemelä-Anttonen, H.; Harra, J.; Koivuluoto, H.; Teisala, H.; Lahti, J.; Tuominen, M.; Kuusipalo, J.; Vuoristo, P.; Mäkelä, J. M. Achieving a Slippery, Liquid-Infused Porous Surface with Anti-Icing Properties by Direct Deposition of Flame Synthesized Aerosol Nanoparticles on a Thermally Fragile Substrate. *Appl. Phys. Lett.* **2017**, *110*, 161603.

(32) Howell, C.; Vu, T. L.; Lin, J. J.; Kolle, S.; Juthani, N.; Watson, E.; Weaver, J. C.; Alvarenga, J.; Aizenberg, J. Self-Replenishing Vascularized Fouling-Release Surfaces. *ACS Appl. Mater. Interfaces* **2014**, *6*, 13299–13307.

(33) Damle, V. G.; Uppal, A.; Sun, X.; Burgin, T. P.; Rykaczewski, K. Rapid and Scalable Lubrication and Replenishment of Liquidinfused Materials. *Surf. Innov.* **2015**, *4*, 102–108.

(34) Cui, J.; Daniel, D.; Grinthal, A.; Lin, K.; Aizenberg, J. Dynamic Polymer Systems with Self-Regulated Secretion for the Control of Surface Properties and Material Healing. *Nat. Mater.* **2015**, *14*, 790–795.

(35) Tenjimbayashi, M.; Nishioka, S.; Kobayashi, Y.; Kawase, K.; Li, J.; Abe, J.; Shiratori, S. A Lubricant-Sandwiched Coating with Long-Term Stable Anticorrosion Performance. *Langmuir* **2018**, *34*, 1386–1393.

(36) Pop, E.; Varshney, V.; Roy, A. K. Thermal Properties of Graphene: Fundamentals and Applications. *MRS Bull.* **2012**, *37*, 1273–1281.

(37) Raj, R.; Maroo, S. C.; Wang, E. N. Wettability of Graphene. *Nano Lett.* **2013**, *13*, 1509–1515.

(38) Hong, G.; Han, Y.; Schutzius, T. M.; Wang, Y.; Pan, Y.; Hu, M.; Jie, J.; Sharma, C. S.; Müller, U.; Poulikakos, D. On the Mechanism of Hydrophilicity of Graphene. *Nano Lett.* **2016**, *16*, 4447–4453.

(39) Shin, Y. J.; Wang, Y.; Huang, H.; Kalon, G.; Wee, A. T. S.; Shen, Z.; Bhatia, C. S.; Yang, H. Surface-Energy Engineering of Graphene. *Langmuir* **2010**, *26*, 3798–3802.

(40) Zang, J.; Ryu, S.; Pugno, N.; Wang, Q.; Tu, Q.; Buehler, M. J.; Zhao, X. Multifunctionality and Control of the Crumpling and Unfolding of Large-Area Graphene. *Nat. Mater.* **2013**, *12*, 321–325.

(41) Ci, H.; Ren, H.; Qi, Y.; Chen, X.; Chen, Z.; Zhang, J.; Zhang, Y.; Liu, Z. 6-Inch Uniform Vertically-Oriented Graphene on Soda-Lime Glass for Photothermal Applications. *Nano Res.* **2018**, *11*, 3106–3115.

(42) Chen, J.; Bo, Z.; Lu, G. Vertically-Oriented Graphene. Springer. **2015**, 978-3-319-15302-5.

(43) Cai, M.; Outlaw, R. A.; Quinlan, R. A.; Premathilake, D.; Butler, S. M.; Miller, J. R. Fast Response, Vertically Oriented Graphene Nanosheet Electric Double Layer Capacitors Synthesized from $C_2H_2$. *ACS Nano* **2014**, *8*, 5873–5882.

(44) Bo, Z.; Mao, S.; Jun Han, Z.; Cen, K.; Chen, J.; Ostrikov, K. Emerging Energy and Environmental Applications of Vertically-Oriented Graphenes. *Chem. Soc. Rev.* **2015**, *44*, 2108–2121.

(45) Shuai, X.; Bo, Z.; Kong, J.; Yan, J.; Cen, K. Wettability of Vertically-Oriented Graphenes with Different Intersheet Distances. *RSC Adv.* **2017**, *7*, 2667–2675.

(46) Rafiee, J.; Mi, X.; Gullapalli, H.; Thomas, A. V.; Yavari, F.; Shi, Y.; Ajayan, P. M.; Koratkar, N. A. Wetting Transparency of Graphene. *Nat. Mater.* **2012**, *11*, 217–222.

(47) Mishra, K. K.; Ghosh, S.; Ravindran, T. R.; Amirthapandian, S.; Kamruddin, M. Thermal

Conductivity and Pressure-Dependent Raman Studies of Vertical Graphene Nanosheets. *J. Phys. Chem. C* **2016**, *120*, 25092–25100.

(48) Preston, D. J.; Lu, Z.; Song, Y.; Zhao, Y.; Wilke, K. L.; Antao, D. S.; Louis, M.; Wang, E. N. Heat Transfer Enhancement during Water and Hydrocarbon Condensation on Lubricant Infused Surfaces. *Sci. Rep.* **2018**, *8*, 1–9.

(49) Ferrari, A. C.; Basko, D. M. Raman Spectroscopy as a Versatile Tool for Studying the Properties of Graphene. *Nat. Nanotechnol.* **2013**, *8*, 235–246.

(50) Heller, E. J.; Yang, Y.; Kocia, L.; Chen, W.; Fang, S.; Borunda, M.; Kaxiras, E. Theory of Graphene Raman Scattering. *ACS Nano* **2016**, *10*, 2803–2818.

(51) Ghosh, S.; Ganesan, K.; Polaki, S. R.; Ravindran, T. R.; Krishna, N. G.; Kamruddin, M.; Tyagi, A. K. Evolution and Defect Analysis of Vertical Graphene Nanosheets. *J. Raman Spectrosc.* **2014**, *45*, 642–649.

(52) Malard, L. M.; Pimenta, M. A.; Dresselhaus, G.; Dresselhaus, M. S. Raman Spectroscopy in Graphene. *Phys. Rep.* **2009**, *473*, 51–87.

(53) Ferrante, C.; Virga, A.; Benfatto, L.; Martinati, M.; De Fazio, D.; Sassi, U.; Fasolato, C.; Ott, A. K.; Postorino, P.; Yoon, D.; Cerullo, G.; Mauri, F.; Ferrari, A. C.; Scopigno, T. Raman Spectroscopy of Graphene under Ultrafast Laser Excitation. *Nat. Commun.* **2018**, *9*, 1–8.

(54) Baranov, O.; Levchenko, I.; Xu, S.; Lim, J. W. M.; Cvelbar, U.; Bazaka, K. Formation of Vertically Oriented Graphenes: What Are the Key Drivers of Growth? *2D Mater.* **2018**, *5*, 044002.

(55) Tzouvadaki, I.; Aliakbarinodehi, N.; Dávila Pineda, D.; De Micheli, G.; Carrara, S. Graphene Nanowalls for High-Performance Chemotherapeutic Drug Sensing and Anti-Fouling Properties. *Sensors Actuators, B Chem.* **2018**, *262*, 395–403.

(56) DuPont Krytox ™ ®. *USA* **2002**, *5*, 8–9.

(57) Kim, P.; Kreder, M. J.; Alvarenga, J.; Aizenberg, J. Hierarchical or Not? Effect of the Length Scale and Hierarchy of the Surface Roughness on Omniphobicity of Lubricant-Infused Substrates. *Nano Lett.* **2013**, *13*, 1793–1799.

(58) Liu, M. S.; Lin, M. C. C.; Wang, C. C. Enhancements of Thermal Conductivities with Cu, CuO, and Carbon Nanotube Nanofluids and Application of MWNT/Water Nanofluid on a Water Chiller System. *Nanoscale Res. Lett.* **2011**, *6*, 1–13.

(59) Cho, J. H.; Gorman, J. J.; Na, S. R.; Cullinan, M. Growth of Monolayer Graphene on Nanoscale Copper-Nickel Alloy Thin Films. *Carbon N. Y.* **2017**, *115*, 441–448.

(60) Enright, R.; Miljkovic, N.; Dou, N.; Nam, Y.; Wang, E. N. Condensation on Superhydrophobic Copper Oxide Nanostructures. *J. Heat Transfer* **2013**, *135*, 091304.

(61) Kwak, K.; Kim, C. Viscosity And Thermal Conductivity Of Copper Oxide Nanofluid

Dispersed In Ethylene Glycol. *Korea-australia Rheol. J.* **2005**, *17*, 35–40.

**Graphic TOC**

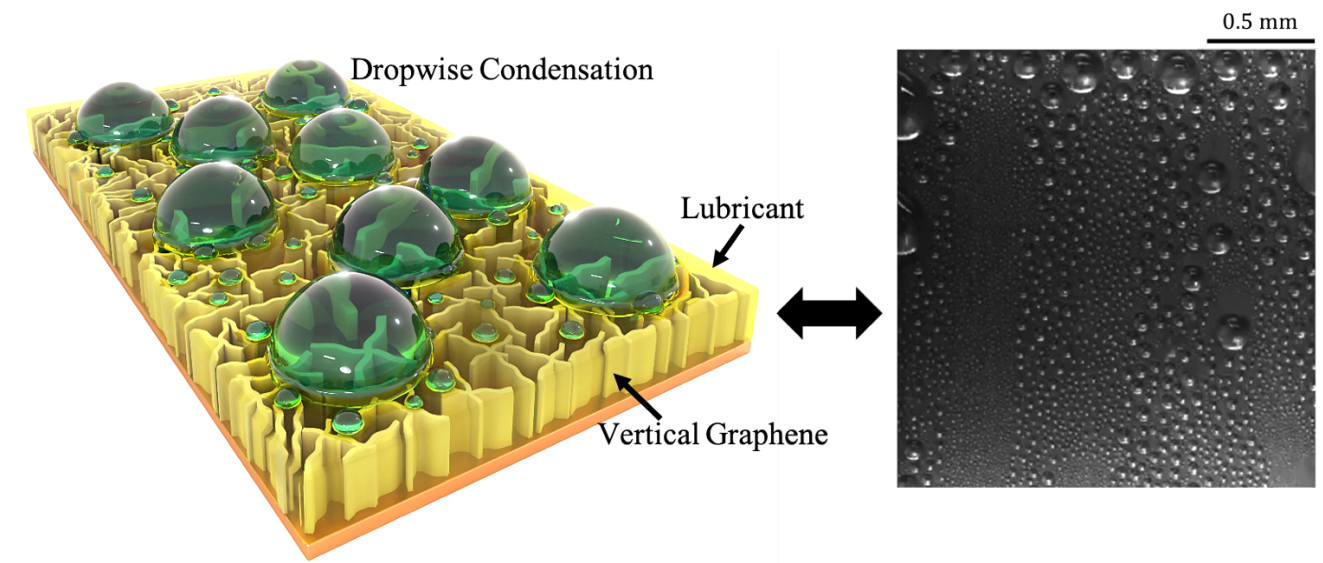

**Supplementary Information for**

# Ultra-Thin Lubricant-Infused Vertical Graphene Nanoscaffolds for High-Performance Dropwise Condensation

Abinash Tripathy[1,+], Cheuk Wing Edmond Lam[1,+], Diana Davila[2], Matteo Donati[1], Athanasios Milionis[1], Chander Shekhar Sharma[3], and Dimos Poulikakos[1*]

[1]Laboratory of Thermodynamics in Emerging Technologies, Department of Mechanical and Process Engineering, ETH Zurich, Sonneggstrasse 3, 8092 Zurich, Switzerland.

[2]IBM Research, Saeumerstrasse 4, 8803 Rueschlikon, Switzerland.

[3]Department of Mechanical Engineering, Indian Institute of Technology, Ropar, Nangal Road, Rupnagar, 140001 Punjab, India.

*Email: dpoulikakos@ethz.ch

[+] These authors have contributed equally to this work.

**Supplementary Figures**

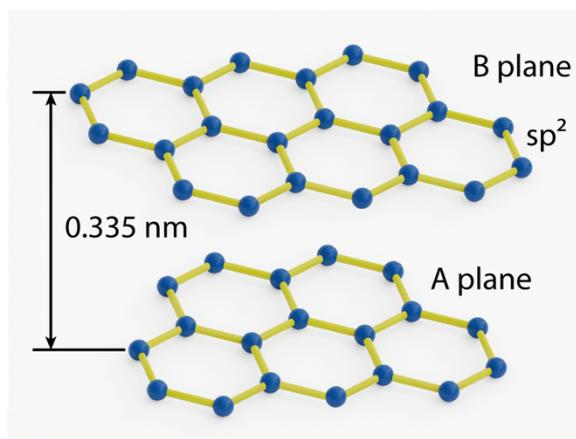

**Supplementary Figure 1:** Schematic of Bernal stacking and the arrangement of carbon atoms in graphene sheets[1]. In-plane strong covalent $sp^2$ bonds exist between the adjacent carbon atoms in graphene resulting in the high in-plane thermal conductivity.

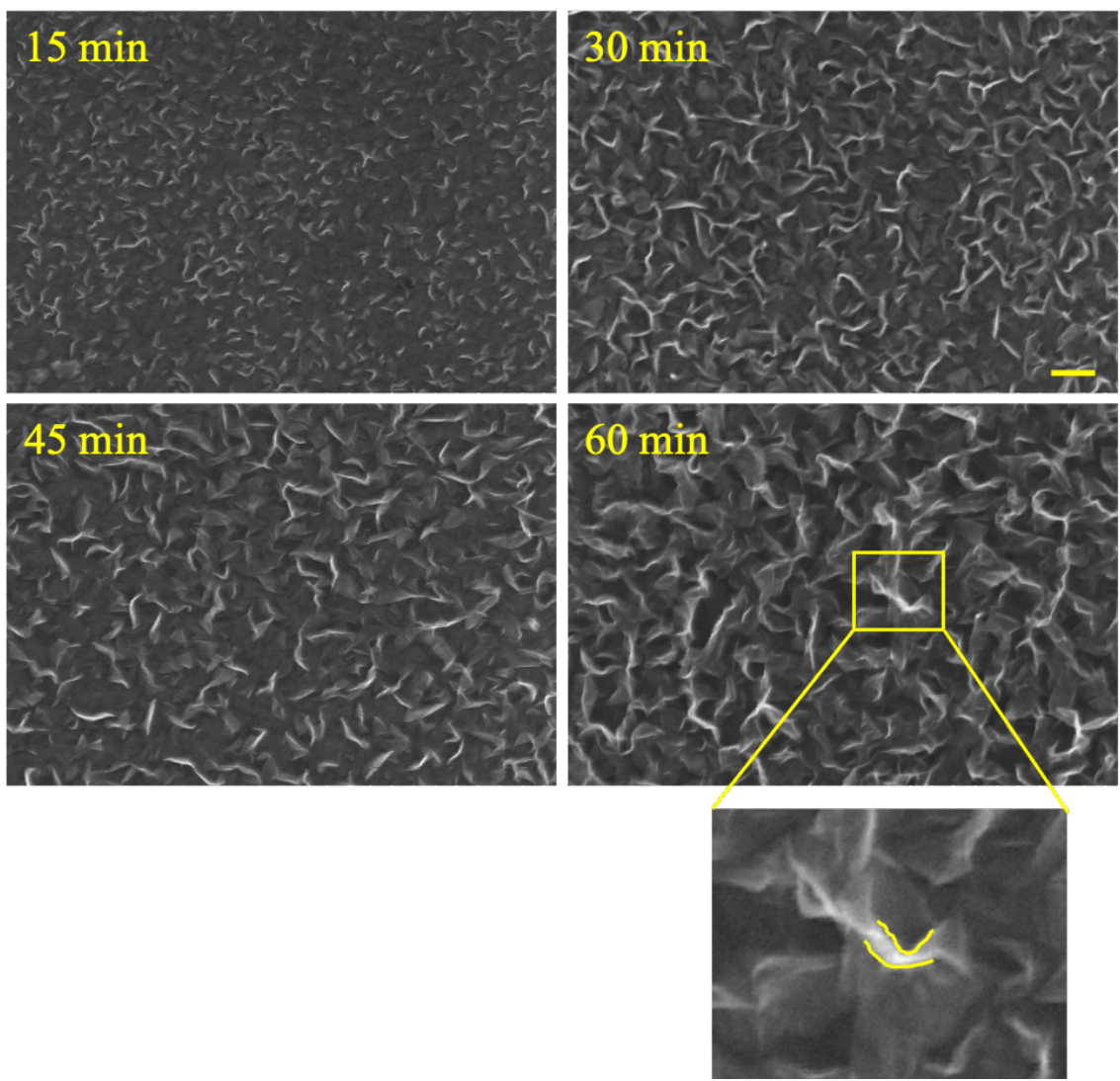

**Supplementary Figure 2:** SEM images of VG samples on copper, grown for different time intervals (Scale bar – 100 nm). Inset showing the edges of the vertical graphene nanosheet.

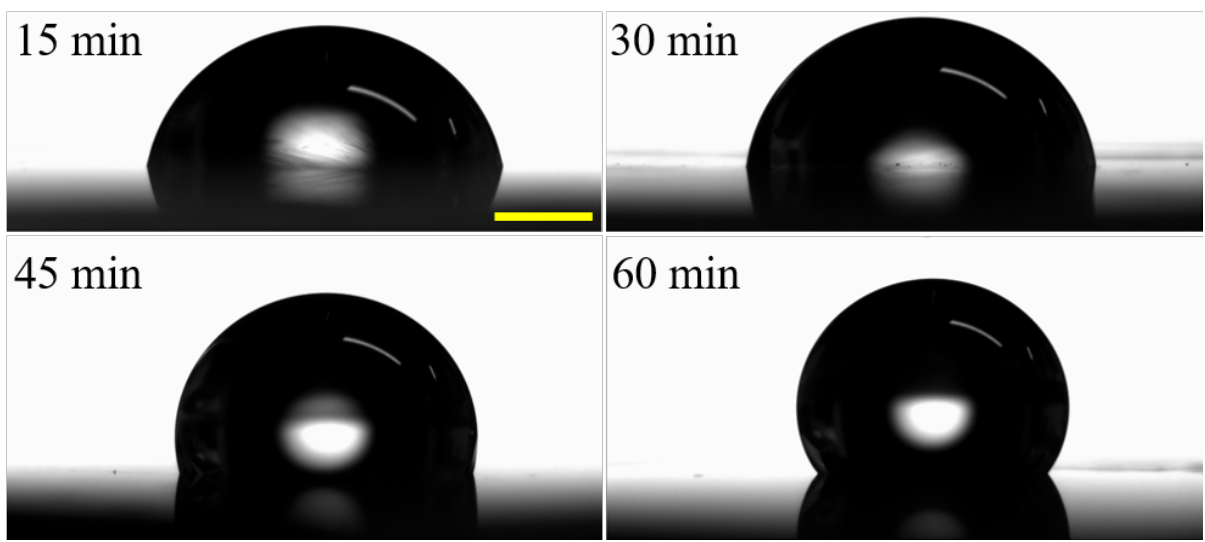

**Supplementary Figure 3:** Water droplets (8 μL) on VG samples with different growth times (Scale bar – 1 mm).

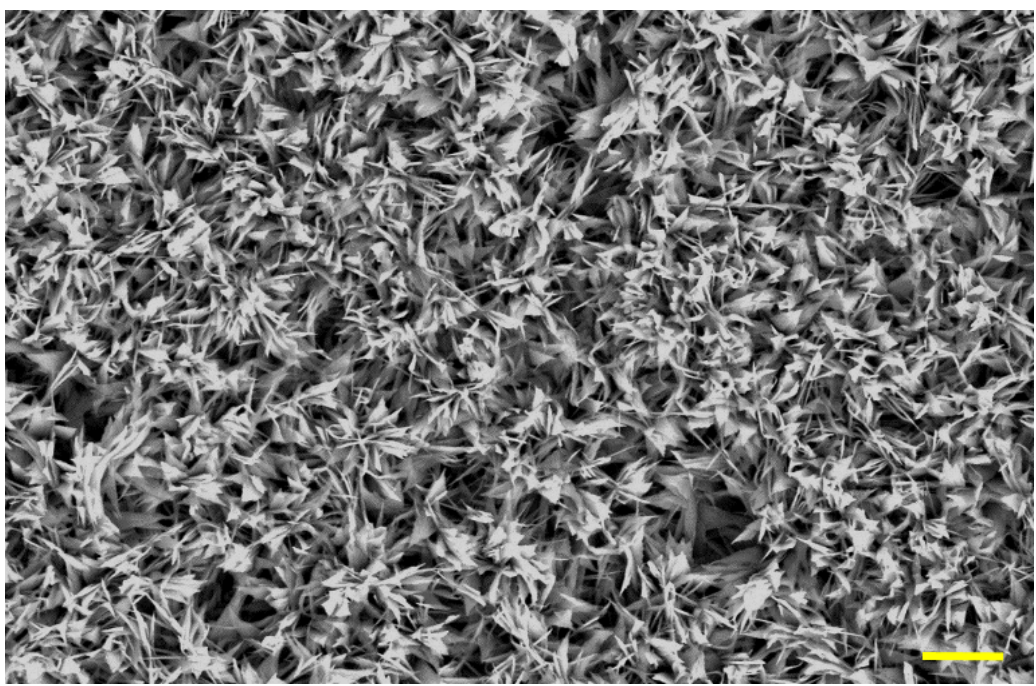

**Supplementary Figure 4:** SEM image of the superhydrophilic CuO nanostructured surface used in the condensation experiments (Scale bar – 2 μm).

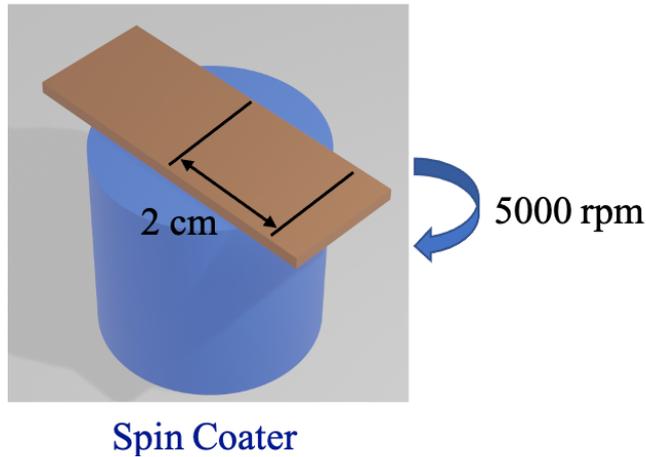

**Supplementary Figure 5:** Schematic of the experimental setup for the spinning test. CAH was measured at a distance of 2 cm from the center of the sample surface after the test.

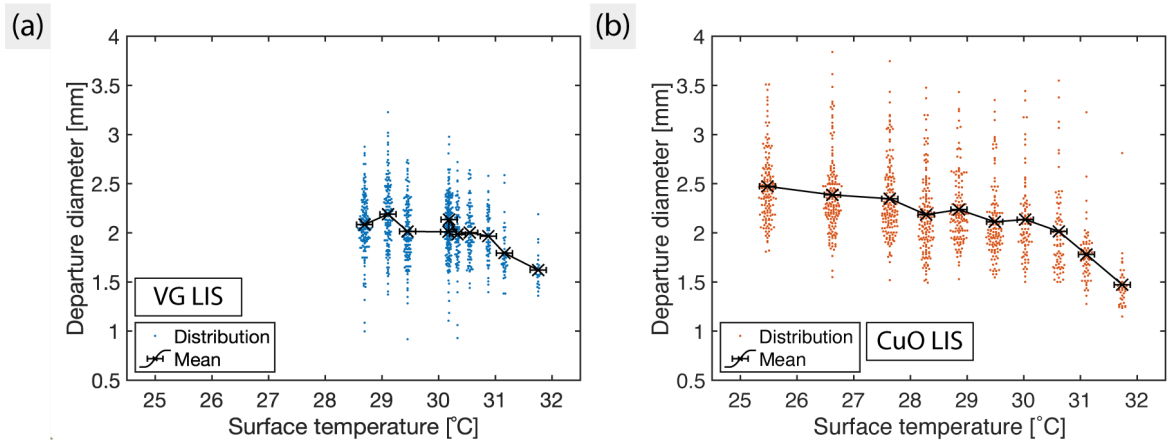

**Supplementary Figure 6:** (a) Distribution of the departure diameters on the VG LIS surface. Each distribution corresponds to specific surface temperature and is arranged in a beeswarm plot centered around that temperature. Each dot in the distribution refers to a droplet measurement. Steam temperature was kept at ~32.9 ˚C. (b) Beeswarm plot for the departure diameters on the CuO LIS surface. Plots generated in MATLAB based on a freely available script[2].

# Supplementary Tables

**Supplementary Table 1:** Comparison of VG LIS with the current state-of-the-art collection of LIS.

| Authors/Journal | Base material for LIS | Lubricant used | Height of micro/nano structures for LIS |
|---|---|---|---|
| Ge et al. *ACS Appl. Mater. Interfaces* 2020, 12, 22246−22255 | CuO, Etched Aluminum ($FeCl_3$ etching) | Krytox1506, ionic liquid: 1-butyl-3-methylimidazolium bis(trifluoromethylsulfonyl)-imide (BMIm), carnation oil, and dodecane. | 20-50 $\mu m$ (Etched Aluminum)  1 $\mu m$ (CuO) |
| Anand et al. *ACS Nano* 2012, 6, 11, 10122–10129 | Silicon micro-post | Krytox, BMIm | 10 $\mu m$ |
| Tenjimbayashi et al. *Langmuir* 2018, 34, 1386−1393 | Self-standing porous surface prepared from Poly(vinylidene fluoride-co-hexafluoropropylene) and Dibutyl phthalate | Perfluoropolyether | 4.24 $\mu m$ |
| Kim et al. *ACS Nano* 2012, 6, 8, 6569–6577 | Electro-deposited polypyrrole | Krytox 100 | 3-4 $\mu m$ |
| Adera et al. *ACS Nano* 2020, 14, 7, 8024–8035 | Copper nanoflorets | Krytox (GPL 104-107) and mineral oil (Hydrobrite 380 PO, Sonneborn) | 3 $\mu m$ |
| Sun et al. *Soft Matter,* 2019, 15, 4808 | $SiO_2$ nanoparticles | Krytox GPL 102 Krytox GPL 104 Krytox GPL 105 Krytox GPL 106 Vacuum pump oil | 1.5 $\mu m$ |
| Sett et al. *Nano Lett.* 2019, 19, 5287−5296 | CuO | Krytox −1525 Krytox −16256 Fomblin − Y25/6 | 1 $\mu m$ |
| Preston et al. *Scientific Reports*, 540 (2018) | CuO | Krytox GPL 101 | 1 $\mu m$ |
| Xiao et al. | CuO | Krytox GPL 101 | 1 $\mu m$ |
| *Scientific Reports*, 3, 1988 (2013) | | | |
| Juuti et al. *Applied Physics Letters* 110, 161603 (2017) | $TiO_2$ nanoparticles | silicone oil with a viscosity of 50 cSt | 0.5 $\mu m$ - 1 $\mu m$ |
| Tesler et al. *Nature Communications*, 6, 8649 (2015) | Nanoporous tungsten oxide (TO) films. | Krytox GPL-K103 | 600 nm |
| Weisensee et al. *International Journal of Heat and Mass Transfer* 109 (2017) 187–199 | $Al_2O_3$ nanostructured surface | Krytox 16256 Krytox GPL 106 Krytox 1514 Krytox GPL 100 Carnation Mineral Oil | 500 nm |
| Tripathy, Lam et al. (this work) | Vertical Graphene | Krytox 1525 | 70 nm |

**Supplementary Table 2:** Properties of VG. Thickness reported as the mean of 50 measurements at different locations).

| VG growth time [minutes] | Sidewall thickness [nm] | Advancing contact angle (ACA) [°] | Contact angle hysteresis (CAH) [°] |
|---|---|---|---|
| 15 | ~13 | 80 $\pm$ 1 | 53 $\pm$ 5.5 |
| 30 | ~16 | 82.7 $\pm$ 2 | 71 $\pm$ 9.4 |
| 45 | ~23 | 102.3 $\pm$ 2 | 69 $\pm$ 9.8 |
| 60 | ~32 | 123 $\pm$ 2 | 66 $\pm$ 2.5 |

**Supplementary Table 3:** Advancing contact angle (ACA) and contact angle hysteresis (CAH) of lubricant-infused surfaces (a) before and (b) after the condensation test in Figure 3 of the main text.

(a)

| Sample | ACA [°] | CAH [°] |
|---|---|---|
| VG LIS | 119 $\pm$ 1 | 2 $\pm$ 1 |
| CuO LIS | 119.7 $\pm$ 0.6 | 3.7 $\pm$ 1.5 |

(b)

| Sample | ACA [°] | CAH [°] |
|---|---|---|
| VG LIS | 118.4 $\pm$ 0.6 | 5.4 $\pm$ 0.6 |
| CuO LIS | 120 $\pm$ 1.7 | 11.3 $\pm$ 2.5 |

## Supplementary Sections

**Supplementary Section 1:** Experimental setup for condensation heat transfer measurement

We used an in-house experimental setup for the condensation tests similar to our recent work[3]. Condensation tests were performed with saturated steam at a nominal pressure of 50 mbar (corresponding saturation temperature 32.87 ˚C). The setup is described in the following whereas the operating procedures are detailed in Supplementary Section 2.

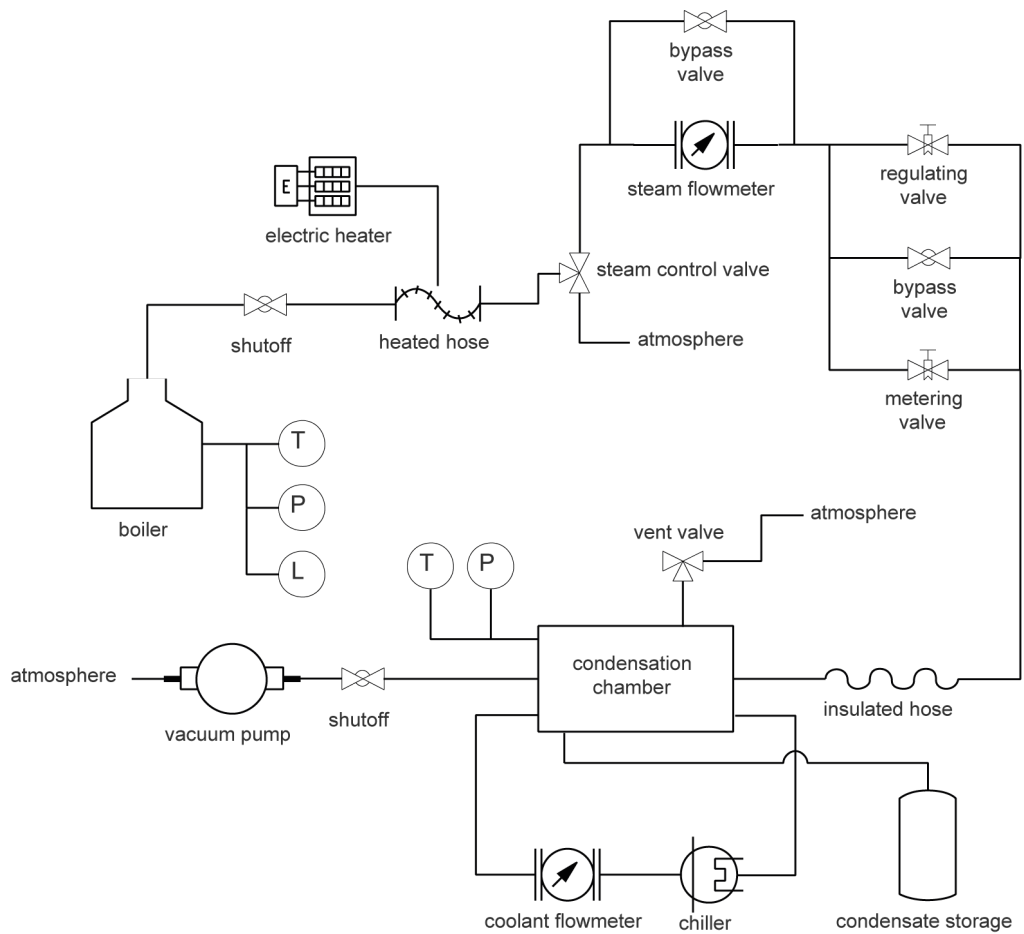

**Figure S1.1:** Schematic of overall setup, with some modifications from our recent work[3].

Figure S1.1 shows a schematic of our overall setup. A boiler was used to generate steam at 1.4 ± 0.01 bar from deionized water. The steam passed through a heated hose to a 3-way valve, where one side was open to atmosphere and the other side was connected to the condensation chamber. The side open to atmosphere was used to degas the deionized water as described later. To the chamber side, the steam would pass through a flowmeter (FAM3255, ABB) arranged in parallel to a bypass valve. A regulating valve (SS-6BMRG-MM, Swagelok), a metering valve (SS-6BMW-MM, Swagelok) and a bypass valve were used to control the flow into the chamber to achieve the target pressure (50 mbar). The steam would then pass horizontally over the test section, where the sample was vertically mounted on a cooled surface of the heat sink. Steam pressure and temperature, as well as the sample temperature were measured in this section. The liquid condensate was collected with a drain at the bottom of the condensation chamber. The steam exited the chamber to a vacuum pump (RC 6, VACUUBRAND) running continuously throughout the experiment.

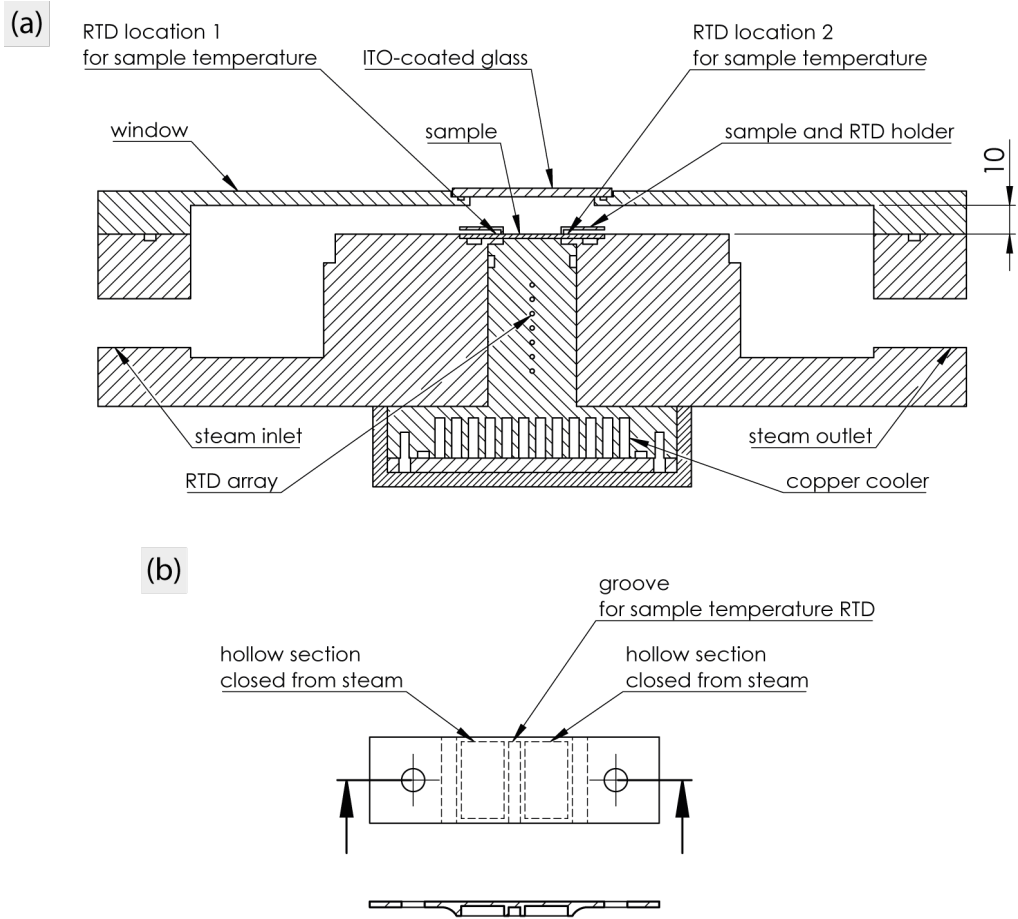

**Figure S1.2:** Schematic of the condensation chamber, with some modifications from our recent work[3], shown without O-rings. (a) Top-view cross section of the test section. Steam entered from the left and exited from the right. Sample temperature RTDs were protected from steam by polycarbonate holders, which held the sample as well. (b) Top view (top) and side-view cross section (bottom) of the polycarbonate holder. The hollow holder protected the sample sides from direct exposure to steam, while minimizing conduction heat transfer through the holder. The groove allowed placement of sample temperature RTDs.

A cross-sectional view of the condensation chamber from the top is depicted in Figure S1.2a. The chamber consisted of a main body milled out of a block of polyether ether ketone (PEEK) and a window milled out of a plate of poly(methyl methacrylate) (PMMA). In operation, the steam entered from the left and exited from the right. A 50 mm × 20 mm sample was placed on a 20 mm × 20 mm cooled surface and the center 20 mm × 20 mm of the sample was exposed to the steam. The 2 sides of size 15 mm × 20 mm were insulated from direct steam exposure by 3D-printed polycarbonate hollow mounts, under which the 2 Pt 1000 Class A sample temperature resistance temperature detectors (RTDs) (P1K0.516.1K.A.152.D.S, IST) were attached onto the sample with Kapton tape. The mean of the 2 sensors was taken as the measured value. Chamber pressure was monitored with a capacitance gauge (CMR 362, Pfeiffer

Vacuum). Chamber temperature was monitored with 2 Pt 1000 Class A RTDs (P1K0.516.1K.A.152.D.S, IST) and the mean of the 2 sensors was taken as the measured value.

The sample with applied thermal paste of thermal conductivity 9.2 W m$^{-1}$ K$^{-1}$ (KP 99, Kerafol) on its back was placed on the cooled surface and a pressure was maintained by the polycarbonate mounts screwed into the chamber to enhance thermal contact. The cooled surface was the front of a heat sink milled out of a copper block (CW004A, Durofer), the opposite of which consisted of a heat exchanger with a coolant recirculated with a chiller (WKL 2200, LAUDA). Between the cooled surface and the heat exchanger, 7 Pt 100 Class A RTDs (Thermo Sensor) were placed as an array to measure the thermal gradient. With the thermal conductivity of the copper heat sink known (394 W m$^{-1}$ K$^{-1}$), the heat flux through the exposed sample surface could be computed. The coolant flow rate was monitored with a flowmeter (SITRANS FM MAG5000 and SITRANS FM MAG 1100, SIEMENS).

A borosilicate glass plate coated with a thin layer of indium tin oxide (Diamond Coatings) was installed on the PMMA window for a heated viewport to remove fog as necessary.

All sensors of the experimental setup were connected to a data acquisition unit (Beckhoff) and the quantities were measured and recorded at 2 Hz using LabVIEW.

A DSLR (D7500, Nikon) with a macro lens of minimum optical distortion (AF Micro-Nikkor 200mm f/4D IF-ED, Nikon) was used to observe and record the condensation behavior. Videos were captured for HTC measurement experiments (Figure 3 in the main text) directly from the camera with an HDMI adapter (Cam Link 4K, elgato) and open-source software OBS. For the durability tests, photographs were taken at every 20 s using the internal timer of the camera.

**Supplementary Section 2:** Operating procedures of the condensation experimental setup

All condensation tests followed the same procedure to ensure consistency. The procedures were similar to our recent work[3].

Before each experiment, the chamber was evacuated overnight with the vacuum pump to allow all water condensate from previous experiments and other volatile substances escape.

The sample was then mounted onto the heat sink and the chamber was re-evacuated. The pump was left to operate continuously until the end of the experiment.

The boiler was filled with deionized water to exceed a predetermined minimum water level. Before steam was introduced into the evacuated chamber, the water was boiled at > 1.4 bar for 30 minutes while open to atmosphere to expel and reduce dissolved gases from the liquid. Meanwhile, the recirculating chiller was set to 37 °C to ensure that when steam was introduced into the chamber at 50 mbar no condensation would immediately occur (steam saturation temperature 32.87 °C).

After that, the bypass valves were opened and closed repeatedly to remove trapped gases in the valves. Once the vacuum stabilized, the bypass valves were closed and the 3-way valve before the flowmeter was switched to the chamber side. The chamber pressure slowly increased by controlling the regulating and metering valves. As the pressure approached 50 mbar, the chiller was set to the first set point of 30 °C. Initial condensation would then be seen as the sample temperature decreased to below the steam temperature.

As the chiller reached a set point, the coolant flow rate was set to $180 \pm 10$ L h$^{-1}$. Boiler power and valves were adjusted to obtain a steady state. The steady state was kept for 1 minute without modifying any settings as 120 data points for each quantity were measured and recorded at 2 Hz by the data acquisition system. During the steady state, the boiler pressure had to maintain at $1.4 \pm 0.01$ bar and the chamber pressure had to maintain at $50 \pm 0.5$ mbar.

The chiller was then set to proceed to the next set point, 25 °C. For the heat transfer coefficient measurements, 10 set points were performed, from 30 °C to -15 °C at 5 °C intervals. Upon arriving at each set point, the coolant flow rate was set to 180 L h$^{-1}$ to compensate for thermal contraction of the lubricant. These 10 set points would then translate to 10 different subcooling values for the samples.

As the steam temperature was higher than the ambient temperature of the laboratory (~ 23 °C), there was fogging on the window. As soon as sufficient fog built up to obstruct a clear view, 10 V was applied for 15 s to the indium-tin-oxide-coated glass plate, resulting in a measured input power of approximately 6.1 W, to heat and remove the fog.

At the end of the experiment, i.e., after the last set point was reached, the chiller was set to heat up to 25 °C. The boiler was turned off, but steam would still pass over the sample as the pump was continuously operating. Once the coolant temperature was sufficiently higher than the atmospheric freezing point (~ 5 °C), the steam supply as well as the pump connection to the chamber were cut off. The chamber was then vented. The sample was then removed and dried with a weak N$_2$ flow.

The full procedure from steam introduction into the chamber to the removal of the sample would typically last around 2 hours.

From the first to the last set point of the experiment, the DSLR was used to capture videos at a 1/60 s shutter speed, with an aperture of f/4.8 and ISO 3200. Resolution was 3840 × 2160. No additional lighting was provided except ambient light (fluorescent lamps) in the laboratory to avoid undesirable heating of the sensors from strong light sources. Videos were captured at lossless quality, which were then transcoded to H.265 using the ffmpeg libx265 software encoder and imported into Adobe Premiere Pro for post-processing.

For durability tests, all procedures remained the same except there was no change of chiller set point for the entire 7 h period. The chiller set point was reduced from 37 ˚C to a set temperature to achieve the desired surface temperature and was kept constant for 7 hours. During this period, the pump had to be occasionally stopped briefly for maintenance. In these events, the connection to the pump was temporarily closed and the steam flow was reduced by adjusting the regulating valve. After maintenance, the original valve setting was restored, and the pump was connected to the chamber again. Such maintenance would last for only a few minutes.

The DSLR was set to capture photographs in RAW format every 20 s using the internal timer for the full duration of the durability test, at a 1/30 s shutter speed, with an aperture of f/5 and ISO 3200. Resolution was 5568 × 3712. Similarly, no additional lighting was provided except ambient light in the laboratory.

**Supplementary Section 3:** Leakage rate of the condensation chamber

Leakage of the chamber would introduce non-condensable gases, which would reduce heat transfer improvement in dropwise condensation. To estimate the leakage rate of our chamber at operating conditions, we performed leakage tests of our condensation chamber.

We evacuated the chamber overnight with the vacuum pump to allow all water condensate from previous experiments and volatile substances to escape. Just before the test, the connection to the pump was opened and closed repeatedly to remove trapped gases in the ball valve.

The valve was subsequently closed. As the chamber pressure stabilized, the leakage test began. We first obtained the initial leakage rate at ~ 1 mbar, as seen in Figure S3.1a. The average leakage rate measured was 0.41 mbar h$^{-1}$ by linear fitting.

Next, we vented the chamber with ambient air to ~ 48 mbar to obtain the leakage rate at close to the typical operating pressure of 50 mbar, as seen in Figure S3.1b. Note that this test was much longer, and the chamber temperature varied over time due to laboratory air conditioning, which affected the pressure measurements by thermal expansion. We thus performed the test for 24 h to compensate for the 24-h air conditioning cycle and thus obtained an average leakage rate of 0.05 mbar h$^{-1}$ by linear fitting.

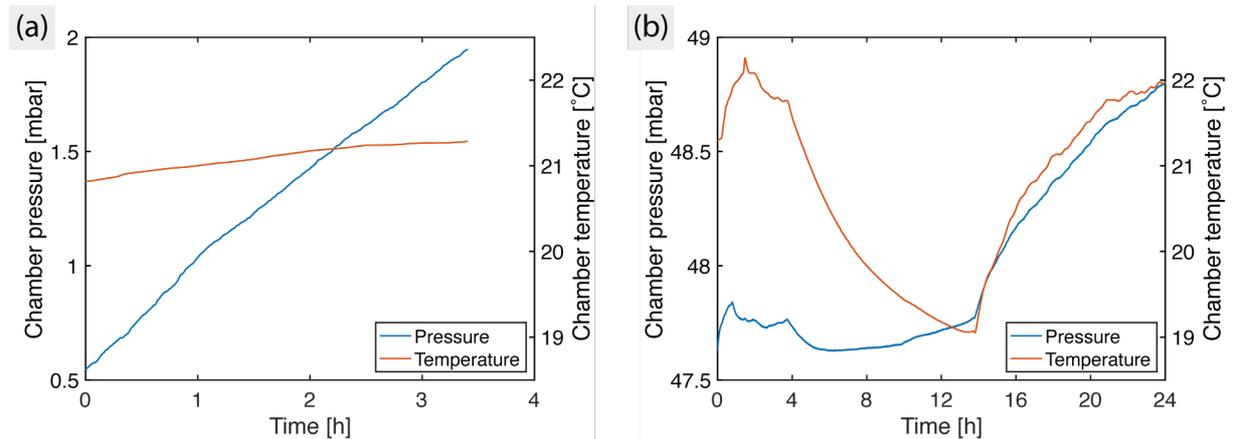

**Figure S3.1:** Chamber leakage tests at (a) ~ 1 mbar for ~ 3.5 h and (b) ~ 48 mbar for 24 h. Data obtained at 2 Hz and presented as 1-minute moving average. y-axes presented at the same scale for (a) and (b).

We chose air as the working fluid for this test as 50 mbar saturated steam has a higher temperature than ambient and heat flow out of the chamber would inevitably reduce the pressure, resulting in an underestimation of the leakage.

From these results we therefore concluded that the leakage was insignificant for the condensation experiments. Note that in operation, there was as well a continuous supply of fresh steam from the boiler and the influence from inward atmospheric leakage could be considered as negligible.

**Supplementary Section 4:** Steam flow rate and velocity

In the condensation chamber, the flow was driven by a pressure difference. As the pressure source (boiler at 1.4 bar) and sink (power of the vacuum pump) were held constant across experiments, the flow rate would be correlated with the amount of heat transferred out from the system. Cooling and condensation would reduce the pressure, thus required a higher flow rate to sustain the same steam pressure of 50 mbar.

This is evident as seen in the heat flux and steam flow rate measurements (Figure S4.1). We computed the characteristic steam flow velocity in the chamber from the volume flow rate of the steam from the boiler. The boiler outflow volume flow rate was measured with the flowmeter factory-calibrated to 1.4 bar saturated steam (~10% uncertainty). By mass conservation, the volume flow rate at 1.4 bar was then converted to the volume flow rate at 50 mbar using the corresponding saturated steam densities with open-source code CoolProp[9]. The characteristic steam velocity was obtained by dividing the chamber steam inflow volume flow rate by the available flow area close to the exposed sample location.

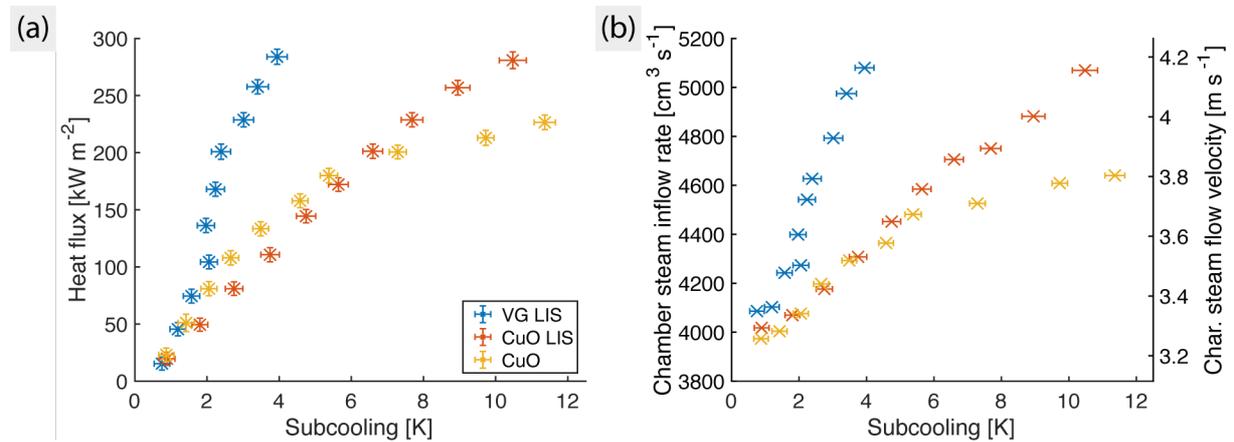

**Figure S4.1:** (a) Heat flux and (b) chamber steam flow speed at different subcooling.

As seen in Figure S4.1, there was a clear correlation between heat transferred out of the chamber through condensation at the interface and the steam flow speed in the chamber. The HTC (and heat flux) measurements in the main text thus correspond to characteristic steam flow velocities of approximately 3.2 – 4.2 m s$^{-1}$. It is worth noting that the corresponding chamber steam inflow rate of 3800 – 5200 cm$^3$ s$^{-1}$ translates to a mass flow rate of only 0.13 – 0.18 g s$^{-1}$, due to the low density of 50 mbar saturated steam (35.48 g m$^{-3}$).

S-13

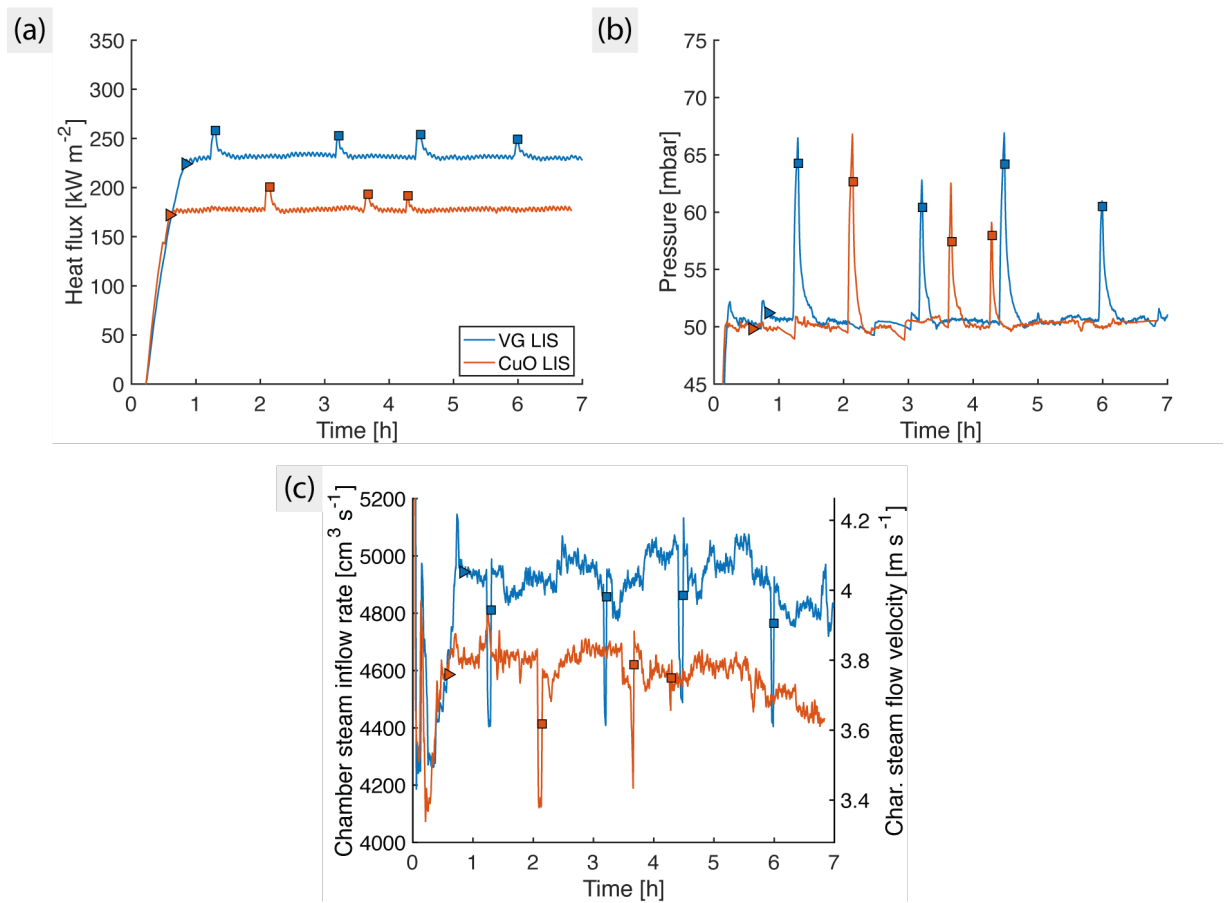

**Figure S4.2:** 1-minute moving average of (a) heat flux, (b) steam pressure, and (c) chamber steam flow speed during the 7-h durability test. Squares indicate when the flow had to be temporarily interrupted for maintenance. Triangles indicate when the heat sink reached the target temperature.

Similar trends were seen in the durability test (Figure S4.2). The flow rate correlated positively with the measured heat flux.

In general, the steam inflow rate was in the range of 4500 – 5000 cm$^3$ s$^{-1}$ at 50 mbar, which translates to a characteristic steam flow velocity of approximately 3.8 – 4.1 m s$^{-1}$ for the durability tests.

**Supplementary Section 5:** Sample mounting and thermal resistance of the lubricant layer

To facilitate a tight attachment of the sample temperature sensors to the sample, a cleanroom towel was used to remove the lubricant on the sides of the sample where steam would not come into direct contact (Figure S1.2).

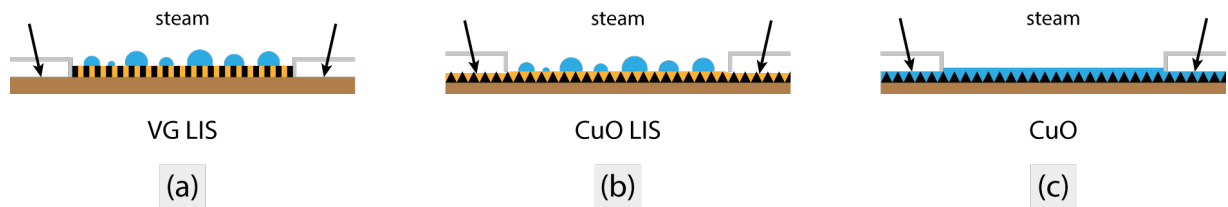

**Figure S5.1:** Location of temperature sensors on the surfaces to measure its temperature during condensation indicated by black arrows. (a) The sensors on VG LIS were placed directly on the copper substrate after removal of the VG and lubricant on the sides. (b) The sensors on CuO LIS were placed on top of the CuO nanostructures after removal of the very top layer of lubricant on the sides. (c) The sensors on the CuO nanostructured surface were placed on top of the nanostructures without treatment. During condensation, water filled the nanostructures and formed a film below the sensor. Schematic not to exact scale. Shapes in brown, yellow and blue indicate copper substrate, lubricant and water respectively. Black rectangles indicate VG. Black triangles indicate CuO nanoblades. Hollow mounts in gray protected the sensors from direct contact with steam.

For VG LIS, wiping removed both the lubricant and the VG. The sample temperature sensors thus measured the temperature below the lubricant layer, i.e., on the copper substrate. See Figure S5.1a.

For CuO LIS, wiping removed the lubricant at the top of the structures only. The sample temperature sensors measured the temperature above the lubricant layer. See Figure S5.1b.

For the CuO nanostructured surface, Kapton tape could adhere well without treatment. The sample temperature sensors measured the temperature above the nanostructures. See Figure S5.1c. Water filled the nanostructures upon condensation.

In our experiments, we aimed to measure the temperature below the lubricant layer $T_{\text{base}}$ so that its thermal resistance can be taken into account in the computation of HTCs.

For VG LIS, this was automatic since the RTDs were directly mounted on the copper substrate.

For CuO LIS and the CuO nanostructured surface, the temperature below the nanostructures was estimated.

We first measured the height of the nanostructures. For the CuO nanostructured surface, a cavity was drilled with focused ion beam (FIB) to reveal a cross section with the nanoblades spanning a height of 2 μm, which was then taken as the amplitude of the height profile.

A cross section of VG is seen in Figure 1c of the main text. The height of structures was taken as 70 nm.

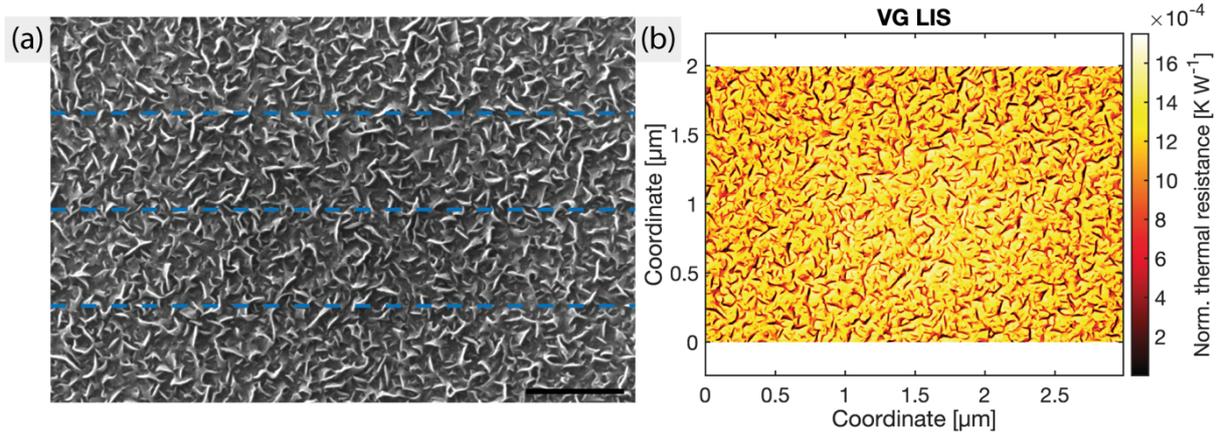

**Figure S5.2:** (a) Top view SEM image of VG without lubricant infused. (Scale bar – 500 nm). Dashed lines refer to the locations for the height profile in Figure S5.5. (b) Estimated thermal resistance distribution of VG LIS, each pixel shown normalized to a surface area of 20 mm x 20 mm in [K W$^{-1}$].

With top-view SEM images of the nanostructures, we estimated their solid and liquid fractions and the resulting thermal resistance. Figure S5.2a is an SEM image of VG before lubricant infusion. The height profile of VG was estimated by assuming a linear relationship between local height and pixel intensity, which is validated afterwards. The global maximum and minimum in pixel intensity were set to correspond to the highest ($z_{surf} = 70$ nm) and lowest point ($z_{base} = 0$) in the height profile respectively. Assuming after lubricant infusion the lubricant surface would be flat and at the same height as the height of the structures (70 nm), we estimated the thermal resistance at each location (pixel) of the surface using the one-dimensional heat conduction equation.

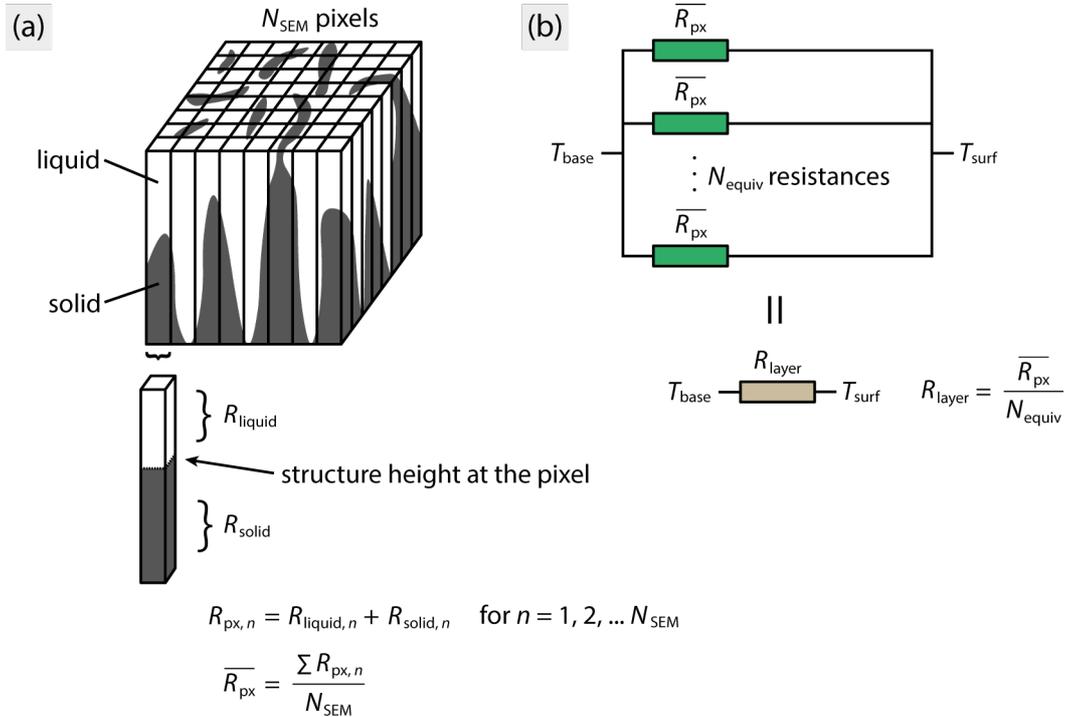

$R_{px,n} = R_{liquid,n} + R_{solid,n}$ for $n = 1, 2, \ldots N_{SEM}$

$$\overline{R_{px}} = \frac{\sum R_{px,n}}{N_{SEM}}$$

**Figure S5.3:** (a) Thermal resistance at a single pixel and averaging over pixels. (b) Parallel thermal resistance circuit for the lubricant layer.

For each pixel of the SEM image, the thermal resistance $R$ [K W$^{-1}$] was defined from the one-dimensional heat conduction equation

$$q = kA \frac{T_{\text{surf}} - T_{\text{base}}}{\Delta z}$$

$$R = \frac{T_{\text{surf}} - T_{\text{base}}}{q} = \frac{\Delta z}{kA}$$

Where $q$ is the heat flow [W], $k$ is the thermal conductivity [W m$^{-1}$ K$^{-1}$], $A$ is the area of conduction [m$^2$] and $\Delta z = z_{\text{surf}} - z_{\text{base}} = z_{\text{surf}}$ is the length of conduction [m]. The resistance for the $n^{\text{th}}$ pixel, $R_{\text{px}, n}$, was computed in a series manner and summed based on the assumption of one-dimensional heat conduction (Figure S5.3a):

$$R_{\text{px}, n} = \frac{\Delta z_{\text{solid}}}{k_{\text{solid}} A_{\text{px}}} + \frac{\Delta z_{\text{liquid}}}{k_{\text{liquid}} A_{\text{px}}}$$

where $A_{\text{px}}$ is the corresponding area of the pixel, dependent on the resolution of the SEM image.

The thermal conductivity of CuO, VG, liquid water, and Krytox 1525 were taken as 33[4–6], 250[7], 0.6[6] and 0.1 (Merck) W m$^{-1}$ K$^{-1}$ respectively.

The thermal resistance of the nanostructure layer $R_{\text{layer}}$ was then calculated from the parallel thermal resistance circuit (Figure S5.3b). We first obtained the average thermal resistance for the area of a pixel, $\overline{R_{\text{px}}}$, by averaging (arithmetic mean) over the pixels in the respective SEM image:

$$\overline{R_{\text{px}}} = \frac{\sum R_{\text{px}, n}}{N_{\text{SEM}}}$$

where $N_{\text{SEM}}$ is the number of pixels available in the SEM image. Then, we computed the equivalent number of pixels, $N_{\text{equiv}}$, for an area of 20 mm x 20 mm, the exposed area in our condensation experiments:

$$N_{\text{equiv}} = \frac{0.02 \times 0.02}{A_{\text{px}}}$$

The thermal resistance of the nanostructure layer $R_{\text{layer}}$ [K W$^{-1}$] for an area of 20 mm x 20 mm could then be obtained with the parallel thermal resistance circuit, for such an area would have $N_{\text{equiv}}$ pixels and thus $\overline{R_{\text{px}}}$ individual resistances arranged in parallel between the lubricant surface temperature $T_{\text{surf}}$ and base temperature $T_{\text{base}}$:

$$R_{\text{layer}} = \frac{1}{\sum \frac{1}{\overline{R_{\text{px}}}}} = \frac{1}{N_{\text{equiv}}/\overline{R_{\text{px}}}} = \frac{\overline{R_{\text{px}}}}{N_{\text{equiv}}}$$

With $R_{\text{layer}}$ of CuO LIS and the CuO nanostructured surface, the subcooling was then computed:

$$\Delta T = T_{\text{steam}} - T_{\text{base}} = T_{\text{steam}} - (T_{\text{surf}} - R_{\text{layer}} q)$$

where $q = 0.02 \times 0.02 \times q''$ is the heat flow through the 20 mm x 20 mm surface.

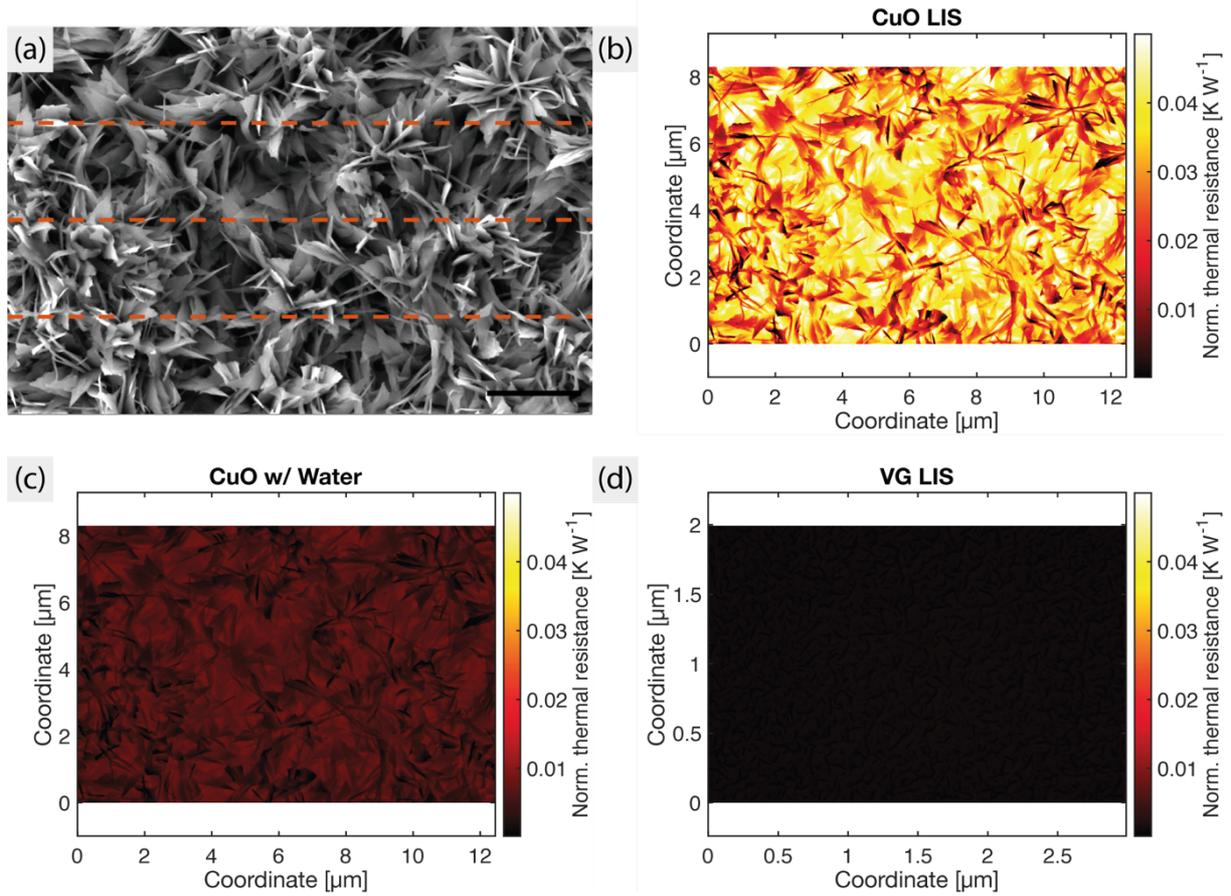

**Figure S5.4:** (a) Top view SEM image of the CuO nanostructured surface without liquids infused. (Scale bar – 2 µm). Dashed lines refer to the locations for the height profile in Figure S5.5. Underlying SEM image reused from our recent work[3] (John Wiley and Sons License 5042641168114). Estimated thermal resistance distribution of (b) CuO LIS, (c) the CuO nanostructured surface infused with water, and (d) VG LIS, at the same color scale. To ensure consistent comparison, the thermal resistance for each pixel is shown normalized to a surface area of 20 mm x 20 mm in [K W$^{-1}$] for each of the 3 surfaces.

We applied the same method to compute the thermal resistance of the nanostructure layer of all 3 surfaces during condensation. Note that upon condensation, water filled up the nanoblades of the CuO nanostructured surface which would otherwise be lubricant when the same structure was used for the fabrication of CuO LIS. Thus, the computation of thermal resistance was performed using their respective thermal conductivities. We present the resistance distribution for VG LIS in Figure S5.2b and compared the resulting resistance distribution on the 3 surfaces with the same color scale in Figure S5.4b to d. As thermal resistance depends on the area (inversely proportional) and the SEM images have different resolutions, their thermal resistances could not be compared pixel-by-pixel. Therefore, we normalized the thermal resistance at each pixel shown in the color maps of Figures S5.2 and S5.4 to a surface area of 20 mm x 20 mm, i.e., $R_{px,\,n}/N_{equiv}$. The thermal resistance of VG LIS was evidently lower than the others (Figure S5.4d). Also, as liquid water has a higher thermal conductivity than the lubricant, the resulting thermal resistance of the nanostructure layer is lower for the CuO nanostructured surface with water than CuO LIS.

S-18

For VG LIS, the thermal resistances of the 5 nm gold coating and the thiol layer were neglected in the computation of the total thermal resistance. For CuO LIS, the thermal resistance of the thiol layer was neglected. These resistances were considered to be negligibly small.

The following table summarizes the nanostructure height, thermal conductivity and estimated thermal resistance:

|  | Nanostructure height [nm] | Solid material | $k_{solid}$ [W m$^{-1}$ K$^{-1}$] | Liquid material | $k_{liquid}$ [W m$^{-1}$ K$^{-1}$] | $R_{layer}$ [K W$^{-1}$] | $R_{layer}$ normalized to 1 m$^2$ [K W$^{-1}$] |
|---|---|---|---|---|---|---|---|
| CuO | 2000 | CuO | 33 | water | 0.6 | 0.0053 | 2.13E-6 |
| CuO LIS | 2000 | CuO | 33 | Krytox 1525 | 0.1 | 0.0317 | 1.27E-5 |
| VG LIS | 70 | VG | 250 | Krytox 1525 | 0.1 | 0.0011 | 4.53E-7 |

To visualize the vast difference in structure heights, a height profile is seen in Figure S5.5, obtained from the concatenation of 3 profiles obtained from horizontal cross sections on the SEM images of VG (Figure S5.2a) and CuO nanoblades (Figure S5.4a).

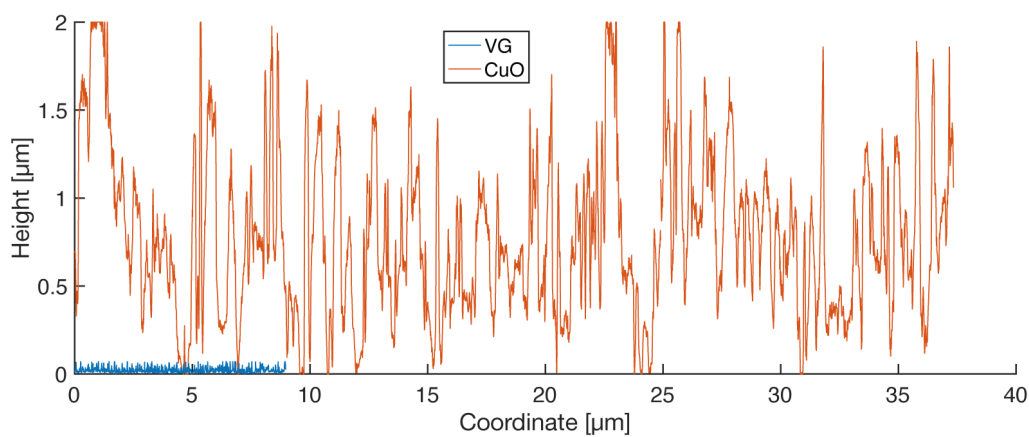

**Figure S5.5:** Height profiles from 3 horizontal lines on the top-view SEM images of VG and CuO.

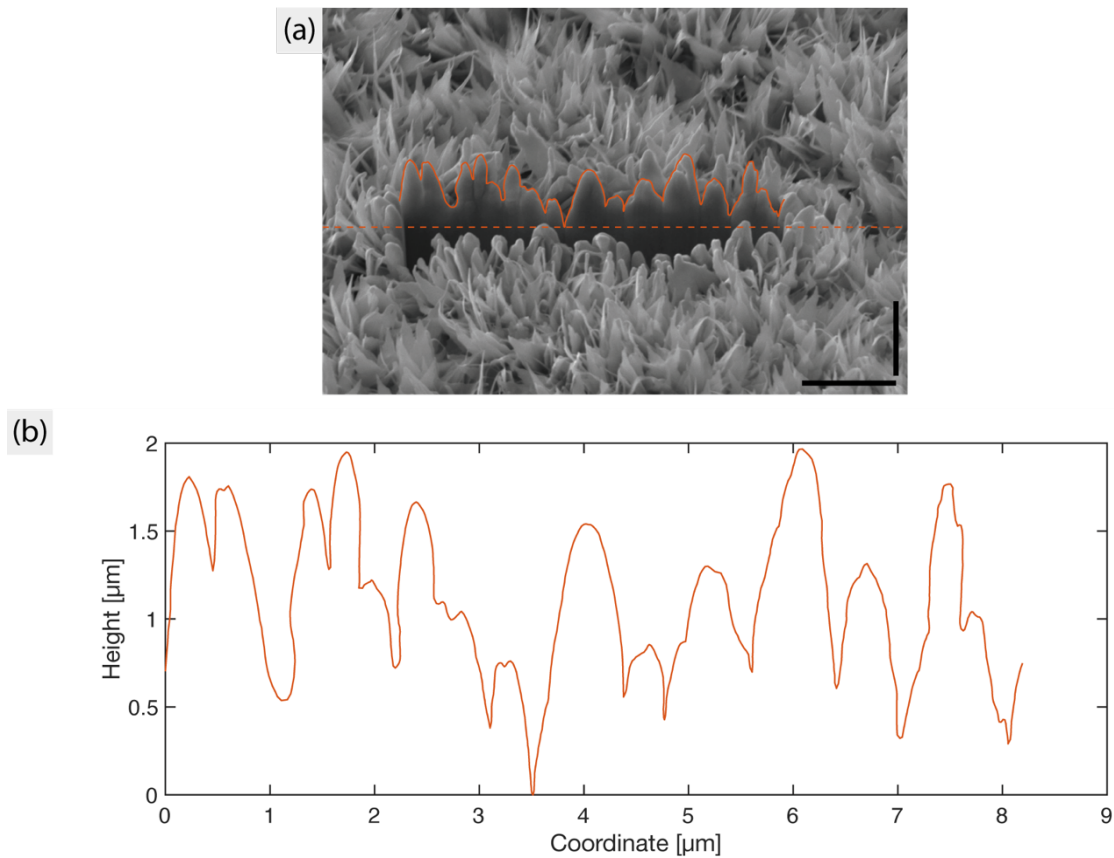

**Figure S5.6:** (a) Height profile extraction from a cross section of the CuO nanostructured surface obtained by FIB drilling. Sample tilted at 53°. (Scale bars – 2 µm, vertical scale bar shorter due to angled view). Underlying SEM image reused from our recent work[3] (John Wiley and Sons License 5042641168114). (b) The obtained height profile. A baseline at $z = 0$ was set at the lowest point of the obtained profile.

We validated the intensity-based method with an angled view of the CuO nanostructured surface. A height profile was manually measured as shown in Figure S5.6a with ImageJ. Choosing a baseline of the structures for $z = 0$ as indicated by the dashed line at the lowest point of the obtained profile, a height profile was extracted, as seen in Figure S5.6b.

The height profile had a mean height of 1.06 µm and a standard deviation of 0.44 µm. The solid fraction was 0.58.

As a comparison, the mean height obtained from the intensity-based 2-dimensional profile in Figure S5.4a was 0.73 µm with a standard deviation of 0.43 µm. The solid fraction was 0.37. We therefore concluded that the assumption of a linear relationship between SEM pixel intensity and the local height, for the estimation of thermal resistances, was reasonable.

For VG, the mean height obtained from the 2-dimensional profile in Figure S5.2a was 24.75 nm with a standard deviation of 14.26 nm. The solid fraction was 0.35.

**Supplementary Section 6:** Computation of heat transfer coefficients and error propagation

For a chiller set point, i.e., a particular subcooling, 120 measurements were taken for 1 minute of steady state (Supplementary Section 2). The mean over these 120 measurements were reported as the measured value. Errors stemming from inherent sensor uncertainties and random fluctuations were presented as error bars. We performed an error analysis of the measured heat transfer coefficient and heat flux values[8], similar to our previous work[3], for the error bars.

HTCs were computed from the heat flux and the subcooling:

$$\tilde{h} = q''/\widetilde{\Delta T} = q''/(T_{\text{steam}} - T_{\text{sample}})$$

In our uncertainty propagation, the estimation of the thermal resistance of the lubricant layer and its influence on the subcooling was not taken into account. We defined a variable $\widetilde{\Delta T} = T_{\text{steam}} - T_{\text{sample}}$, which is the difference between the steam temperature $T_{\text{steam}}$ and the sample temperature $T_{\text{sample}}$, as measured with the RTDs placed in Figure S5.1, resulting in a corresponding HTC $\tilde{h}$.

First the uncertainty in $\widetilde{\Delta T}$ was computed.

The inherent uncertainty of the RTDs used was $\delta_{\text{RTD}} = 0.2$ K. Thus, the steam temperature uncertainty was:

$$\delta_{T_{\text{steam}}} = \frac{1}{2}\sqrt{\delta_{\text{RTD}}^2 + \delta_{\text{RTD}}^2}$$

as the steam temperature was taken as the mean of 2 RTDs.

The sample temperature uncertainty was similar, as it was the mean of 2 RTDs as well:

$$\delta_{T_{\text{sample}}} = \frac{1}{2}\sqrt{\delta_{\text{RTD}}^2 + \delta_{\text{RTD}}^2}$$

The uncertainty in $\widetilde{\Delta T}$ was propagated as follows:

$$\delta_{\widetilde{\Delta T}} = \sqrt{\delta_{T_{\text{steam}}}^2 + \delta_{T_{\text{sample}}}^2}$$

Next, the heat flux uncertainty was propagated. The uncertainty of the linear fitting of the RTD array was as follows:

$$\delta_{\text{fit}} = \delta_{\text{RTD}}\sqrt{\frac{N_{\text{RTD}}}{N_{\text{RTD}}\sum x_{\text{RTD}}^2 - (\sum x_{\text{RTD}})^2}}$$

The uncertainty in [K m$^{-1}$] depended on the number of RTDs in the array $N_{\text{RTD}} = 7$, and the location of the RTDs in the array $x_{\text{RTD}} = 0, 0.005, 0.01, 0.015, 0.02, 0.025, 0.03$ m.

The uncertainty in the linear fitting was then propagated to the uncertainty in the heat flux:

$$\delta_{q''} = k\frac{A_{\text{cross}}}{A_{\text{cond}}}\delta_{\text{fit}}$$

where $k$ and $A_{\text{cross}}$ are the thermal conductivity and the cross-sectional area respectively, for the part of the heat sink where the array of RTDs were located, and $A_{\text{cond}}$ (20 mm x 20 mm) was the area of the condensing surface.

Finally, the uncertainty in $\tilde{h}$ was computed:

$$\delta_{\tilde{h}} = \sqrt{\left[\frac{1}{(\widetilde{\Delta T})}\delta_{q''}\right]^2 + \left[\frac{-q''}{(\widetilde{\Delta T})^2}\delta_{T_{\text{steam}}}\right]^2 + \left[\frac{q''}{(\widetilde{\Delta T})^2}\delta_{T_s}\right]^2}$$

The mean values of $\widetilde{\Delta T}$ and $q''$ over 120 readings of 1 minute were used to compute the uncertainty in $\tilde{h}$.

To take into account the fluctuations of measurements over the 120 readings of the 1-minute steady state, we summed the uncertainties above with the standard deviation of the readings, i.e., $\delta_{\widetilde{\Delta T}} + \sigma_{\widetilde{\Delta T}}$, $\delta_{q''} + \sigma_{q''}$ and $\delta_{\tilde{h}} + \sigma_{\tilde{h}}$, where $\sigma$ refers to the standard deviation. The sums were used as the error bars in the figures.

**Supplementary Section 7:** Measurement of droplet departure diameters

Droplet departure diameters were measured at different subcooling values for CuO LIS and VG LIS. The measurement procedures are described in this section.

Videos were taken during the HTC measurement experiments (Supplementary Sections 1 and 2). They were converted to PNG image sequences with Adobe Premiere Pro for frame-by-frame analyses in ImageJ.

To avoid edge effects of the 20 mm x 20 mm area exposed to steam and pinning sites of the CuO LIS which obstructed departure, a crop window with an area of 10 mm x 10 mm was considered. The top edge of the crop window was set to be 3 mm from the top edge of the exposed area, and the left edge was set to be 5 mm from the left edge of the exposed area.

The droplets were measured with a manual circular fit. Droplets were considered when they were falling significantly quicker than nearby droplets. The size of the droplets was measured when the drop was near the bottom edge of the crop window, and the frame at which the departing droplet had the best circular shape was measured. Droplets were included only when the centroid of the measurement (center of the circle fit) lied between the left and right edges of the crop window.

Since frequency was as well desired, a set of criteria was used to determine whether an event would be counted. First, droplets were only considered if they had a clear sweeping track left as they moved downwards. For droplets sliding along the left and right edges of the crop window, they were counted only if the centroid of the measurement lied within the crop window when the droplets were near its bottom edge.

There were cases where departing droplets coalesced near the bottom edge of the crop window. In such cases, the droplets before coalescence were measured and considered as the departure diameter of individual droplets, although they might not have reached near the bottom edge of the crop window. However, the approximate centroid of the coalesced drop should still be between the left and right edges of the crop window when it was near the bottom edge.

It is noted that there might be droplets which began to depart near the bottom edge. Since they did not have a clear sweeping track and they did not move significantly faster than other droplets, these departure events were not counted.

**Supplementary Section 8:** HTC without taking thermal resistance into account

The heat transfer coefficients based on the surface temperature above the nanostructures $T_{surf}$ are reported in Figure S8.1a as a function of the temperature difference $T_{steam} - T_{surf}$, i.e., $q''/(T_{steam} - T_{surf})$. The corresponding heat flux measurements are reported in Figure S8.1b.

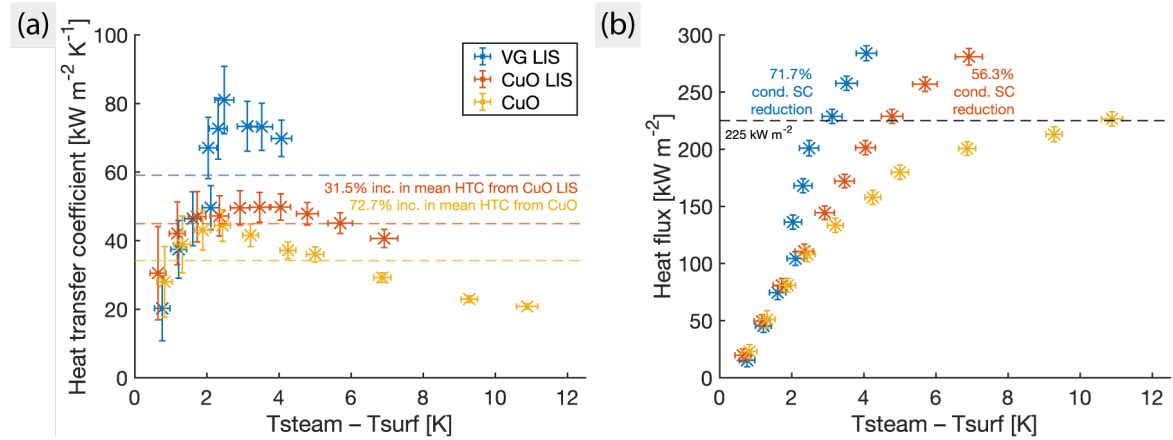

**Figure S8.1:** Heat transfer measurements found in Figure 3a and 3b of the main text, when the lubricant thermal resistances are not taken into account in the computation of HTC. x-axis refers to the difference between steam temperature $T_{steam}$ and surface temperature $T_{surf}$ above the lubricant layer.

The thermal resistance of the lubricant layer is taken out of account from HTC computation. In other words, the thermal resistance computation in Supplementary Section 5 was reversed. Temperature sensors on the CuO nanostructured surface and CuO LIS were readily measuring the temperature above the nanostructures and the infused liquids. For VG LIS, the thermal resistance was used to estimate the temperature on top of the lubricant layer, as during the experiment the temperature sensor was placed directly on the base substrate (Figure S5.1a).

VG LIS presented a 31.5% mean increase in HTC from CuO LIS, and a 72.7% mean increase from the CuO nanostructured surface. At 2.4 K subcooling where VG LIS performed the best, the improvement over the CuO nanostructured surface was 73.8% and that over CuO LIS was 62.6%.

Same trends were observed in heat flux measurements as in the main text. At a heat flux of 225 kW m$^{-2}$, CuO LIS exhibited a 56.3% reduction in the temperature difference $T_{steam} - T_{surf}$ whereas VG LIS exhibited a 71.7% reduction. At 3.8 K subcooling, the heat flux maintained by VG LIS was 84.0% higher than that of the CuO nanostructured surface and 43.3% higher than that of CuO LIS.

**Supplementary Movies**

**Supplementary Movie S1:** Condensation behavior on the CuO nanostructured surface, CuO LIS and VG LIS.

The condensation behavior of the 3 surfaces at a subcooling around 2 K. Playback at actual speed. Droplet pinning sites are seen on the CuO LIS surface. Movie encoded with H.265.

**Supplementary Movie S2:** 7-hour durability test for CuO LIS and VG LIS.

Condensation behavior on CuO LIS and VG LIS for 7 h. Playback at 500x speed. Movie encoded with H.265.

**References**

(1) Pop, E.; Varshney, V.; Roy, A. K. Thermal Properties of Graphene: Fundamentals and Applications. *MRS Bull.* **2012**, *37*, 1273–1281.

(2) Jonas. Plot Spread Points (beeswarm plot). https://ch.mathworks.com/matlabcentral/fileexchange/37105-plot-spread-points-beeswarm-plot. **2021**.

(3) Donati, M.; Lam, C. W. E.; Milionis, A.; Sharma, C. S.; Tripathy, A.; Zendeli, A.; Poulikakos, D. Sprayable Thin and Robust Carbon Nanofiber Composite Coating for Extreme Jumping Dropwise Condensation Performance. *Adv. Mater. Interfaces* **2020**, *8*, 2001176.

(4) Sett, S.; Sokalski, P.; Boyina, K.; Li, L.; Rabbi, K. F.; Auby, H.; Foulkes, T.; Mahvi, A.; Barac, G.; Bolton, L. W.; Miljkovic, N. Stable Dropwise Condensation of Ethanol and Hexane on Rationally Designed Ultrascalable Nanostructured Lubricant-Infused Surfaces. *Nano Lett.* **2019**, *19*, 5287–5296.

(5) Kwak, K.; Kim, C. Viscosity And Thermal Conductivity Of Copper Oxide Nanofluid Dispersed In Ethylene Glycol. *Korea-australia Rheol. J.* **2005**, *17*, 35–40.

(6) Liu, M. S.; Lin, M. C. C.; Wang, C. C. Enhancements of Thermal Conductivities with Cu, CuO, and Carbon Nanotube Nanofluids and Application of MWNT/Water Nanofluid on a Water Chiller System. *Nanoscale Res. Lett.* **2011**, *6*, 1–13.

(7) Mishra, K. K.; Ghosh, S.; Ravindran, T. R.; Amirthapandian, S.; Kamruddin, M. Thermal Conductivity and Pressure-Dependent Raman Studies of Vertical Graphene Nanosheets. *J. Phys. Chem. C* **2016**, *120*, 25092–25100.

(8) Taylor, J. R.; Thompson, W. An Introduction to Error Analysis: The Study of Uncertainties in Physical Measurements. *Physics Today*. **1998**, *51*, 57–58.

(9) Bell, I. H.; Wronski, J.; Quoilin, S.; Lemort, V. Pure and Pseudo-Pure Fluid Thermophysical Property Evaluation and the Open-Source Thermophysical Property Library Coolprop. *Ind. Eng. Chem. Res.* **2014**, *53*, 2498–2508.